\documentclass[9pt,twocolumn,twoside]{pnas-new}

\templatetype{pnasresearcharticle} 

\usepackage{xspace}
\newcommand{\e}{{\langle}e{\rangle}}

\newcommand{\Rstar}{\ensuremath{R_{\star}}\xspace}

\newcommand{\Mp}{\ensuremath{M_p}\xspace}

\newcommand{\Rp}{\ensuremath{R_p}\xspace}

\newcommand{\Me}{\ensuremath{M_{\oplus}}\xspace} 
\renewcommand{\Re}{\ensuremath{R_{\oplus}}\xspace}

\newcommand{\Rsun}{\ensuremath{R_{\odot}}\xspace }
\newcommand{\Msun}{\ensuremath{M_{\odot}}\xspace}

\newcommand{\numplanets}{1646\xspace}
\newcommand{\numstars}{1209\xspace}
\newcommand{\numsingles}{848\xspace}
\newcommand{\nummultis}{798\xspace}
\newcommand{\alderaan}{{\tt ALDERAAN}\xspace}

\usepackage{amsmath}
\usepackage{bm}
\usepackage{mathtools}

\begin{document}

\title{Planets larger than Neptune have elevated eccentricities}

\author[a,1]{Gregory J. Gilbert} 
\author[a]{Erik A. Petigura} 
\author[a]{Paige M. Entrican} 

\affil[a]{University of California, Los Angeles}

\leadauthor{Gilbert}


\significancestatement{The eccentricity (ellipticity) of a planet's orbit is a relic of its formation history. We measured eccentricities of \numplanets planets with sizes ranging from 0.5 to 16 Earth-radii (\Re). On average, large planets (4--16~\Re) are four times more eccentric than small planets (0.5--4~\Re), pointing to distinct formation channels for these two size groups. Small planets typically form on nearly circular orbits and experience minimal perturbations, while large planets are more likely to experience eccentricity excitation. Small planets are bifurcated into at least two groups, super-Earths (1.0--1.5~\Re) and sub-Neptunes (2.0--3.0~\Re), with few planets in between. The planets that fall between these two populations may also have elevated eccentricities, pointing to dynamically exotic formation histories.}

\authorcontributions{G.J.G. led the experimental design, data reduction, statistical analysis, physical interpretation, and manuscript preparation. E.A.P. advised and assisted with all aspects of the project. P.M.E. built custom data visualization software and performed model validation checks.}

\authordeclaration{The authors declare no conflict of interest.}
\correspondingauthor{\textsuperscript{1}To whom correspondence should be addressed. E-mail: gjgilbert@astro.ucla.edu}


\begin{abstract}
NASA's Kepler mission identified over 4000 extrasolar planets that transit (cross in front of) their host stars. This sample has revealed detailed features in the demographics of planet sizes and orbital spacings. However, knowledge of their orbital shapes --- a key tracer of planetary formation and evolution --- remains far more limited. We present measurements of eccentricities for \numplanets Kepler planets, 92\% of which are smaller than Neptune. For all planet sizes, the eccentricity distribution peaks at $\bm{e=0}$ and falls monotonically toward zero at $\bm{e=1}$. As planet size increases, mean population eccentricity rises from $\bm{\e = 0.05 \pm 0.01}$ for small planets to  $\bm{\e = 0.20 \pm 0.03}$ for planets larger than $\sim$~3.5 Earth-radii \Re. The overall planet occurrence rate and planet-metallicity correlation also change abruptly at this size. Taken together, these patterns indicate distinct formation channels for planets above and below $\sim$~3.5~\Re. We also find size dependent associations between eccentricity, host star metallicity, and orbital period. While smaller planets generally have low eccentricities, there are hints of a noteworthy exception: eccentricities are slightly elevated in the ``radius valley,'' a narrow band of low occurrence rate density which separates rocky ``super-Earths'' (1.0--1.5~\Re) from gas-rich ``sub-Neptunes'' (2.0--3.0~\Re). We detect this feature at $2.1\sigma$ significance. Planets in single- and multi-transiting systems exhibit the same size-eccentricity relationship, suggesting they are drawn from the same parent population.
\end{abstract}

\dates{This manuscript was compiled on \today}
\doi{\url{www.pnas.org/cgi/doi/10.1073/pnas.XXXXXXXXXX}}

\maketitle
 \thispagestyle{firststyle}
\ifthenelse{\boolean{shortarticle}}{\ifthenelse{\boolean{singlecolumn}}{\abscontentformatted}{\abscontent}}{}



\dropcap{T}he eight planets of the Solar System have nearly circular orbits, a fact that provided early motivation for the nebular hypothesis of planet formation \citep{Laplace1796}. Their small but non-zero eccentricities (ellipticities) have a mean value  $\e = 0.06$, which has been interpreted as the result of moderate orbital migration and resonance crossing \citep{Tsiganis2005}. The Solar System planets may be grouped according to three size classes: terrestrials ($R_p$ = 0.5--1.0~Earth-radii or \Re), ice-giants (${\sim}4$ \Re), and gas-giants (${\sim}10$~\Re). In contrast, NASA's Kepler mission showed that extrasolar planets span a continuum of sizes and that small planets, with sizes between 1 and 4 \Re and orbital periods less than a year, are ubiquitous and occur at rate of $\sim$1 per star \citep{Petigura2013}. Small planets out-number large planets by an order of magnitude \citep{Howard2012}. Moreover, small planets are bifurcated into two distinct groups: rocky super-Earths (1.0--1.5~\Re) and gas-rich sub-Neptunes (2--3~\Re) with few planets in between \citep{Fulton2017}. While the size distribution of close-in small planets has been characterized to high detail, their eccentricities remain far more uncertain. 

Giant exoplanets are known to possess a wide range of eccentricities, from nearly circular ($e=0$) to highly elliptical (the current record-holder is HD20782b with $e=0.97$, \citealp{OToole2009}). Hot Jupiters, with orbital periods $P < 10$~days and sizes ranging from 1--2$\times$ Jupiter, tend to have small eccentricities, likely the result of tidal circularization \citep{RasioFord1996, Jackson2008, DawsonJohnson2018}. In contrast, longer period Doppler-detected Jovians have a wide range of eccentricities with a mean value of $\e \approx 0.3$ \citep{Kipping2013-beta}. Doppler measurements of small planet eccentricities are often prohibitively expensive since their Doppler amplitudes are much smaller. 

Transit-based analyses have provided some insight into the eccentricities of small planets. In these studies, the uncertainties on individual planet's eccentricities are large ($\sigma_e\sim0.3$ is a typical value; e.g., \citealp{VanEylen2019}). Fortunately, the census of transiting planets is large enough that many imprecise measurements may be combined to yield insights into the population-level distribution of eccentricities. Prior studies have found that planets in single-transiting systems are more eccentric than planets in multis: $e_{\rm single}\sim0.3$ and $e_{\rm multi}\sim0.05$ \citep{Fabrycky2014, Xie2016, Mills19, VanEylen2015, VanEylen2019, SagearBallard2023}. These studies fall into two categories: (1) analyses that used transit durations \citep{Fabrycky2014, Xie2016, Mills19}, and (2) analyses that modeled the full transit profile \cite{VanEylen2015, VanEylen2019, SagearBallard2023}. Duration-based studies are computationally efficient, but lead to a loss of information. Transit profiling involves more human and computational effort, but to date has been restricted to samples of ${\sim}100$ planets. Our present work extends the latter method to over 1000 planets, enabling a more detailed exploration of eccentricity, in particular its relation to other star and planet properties.

The basic paradigm of planet formation by core-nucleated accretion has be remarkably successful at at explaining the observed population of close-in exoplanets \citep{Pollack1996}. However, when, where, and how planet cores grow; accrete and lose gaseous envelopes; and dynamically interact with their siblings remains poorly understood. Many proposed processes imprint themselves on the extant distributions of exoplanet sizes, eccentricities, and other orbital characteristics. For example, planet-planet scattering \citep{FordRasio2008} and Lidov-Kozai oscillations \citep{FabryckyTremain2007} excite eccentricity, whereas tidal damping \citep{Jackson2008}, inelastic mergers \citep{Li2021}, and disk-driven migration \citep{GoldreichSari2003} quench it. Each mechanism predicts distinct relationships between eccentricity $e$, planet radius $R_p$, host star metallicity [Fe/H] (i.e., stellar inventory of elements heavier than helium), and orbital period $P$. The goal of this paper is to uncover these relationships.

\section*{Methods}\label{sec:methods}
We begin with a high-level summary of our methods here before expanding upon them in the sections below. We first select a sample of \numstars single Sun-like stars hosing \numplanets planets (see Figure~\ref{fig:population}). Kepler gathered some 40,000 photometric measurements of each star over its four-year mission. Each transit may be described with five transit shape parameters. We derive these parameters with particular care to preserve covariances and uncertainties, thereby compressing the high-dimensional photometric dataset into a low-dimensional catalog of transit shape measurements. Next, we combine transit shape and stellar density measurements to derive eccentricities of individual planets. Finally, we characterize the eccentricity distributions of various sub-samples of planets with a hierarchical model. The flow from photometry to eccentricities is shown in Figure~\ref{fig:methods}.

\subsection*{Defining the star/planet sample}

From 2009-2013, the Kepler spacecraft continuously monitored ${\sim}150,000$ Sun-like stars, yielding a sample of over 4000 exoplanet candidates with high reliability (false positive rate below $2\%$ for $P<100$ days; \citealp{Thompson2018}) and well-characterized completeness \citep{Christiansen2020}. This enabled the measurement of exoplanet occurrence rates as a function of planet size and period (see \citealp{Bryson2021} and references therein). The Kepler team measured orbital period to exquisite precision, and subsequent efforts have measured planetary sizes to ${\sim}3-10\%$ for the bulk of the sample \citep[e.g.,][]{Fulton2017, Berger2018}.

Starting with the final Kepler catalog of planet candidates \citep{Thompson2018}, we eliminated any objects marked as false positives, leaving 4078 planets orbiting 3087 stars. We cross-matched these stars against a single homogeneous stellar properties catalog based on {\em Gaia} astrometric and photometric measurements \citep{Berger2020-stars} and selected stars with the following properties: radius $R_{\star} = 0.7{-}1.4$~\Rsun, effective temperature $T_{\rm eff} = 4700{-}6500$~K, and surface gravity $\log g > 4.0$. These criteria correspond to main sequence stars with masses $M_{\star} = 0.7-1.3$~\Msun, or mid-K to late-F spectral type. Finally, we eliminated any stars with greater than $5\%$ flux contamination from nearby sources, as identified by \citealp{Furlan2017}, as well as any stars with a Gaia renormalized unit weigh error RUWE $> 1.4$; such objects have a high probability of being unresolved binaries with separations between 0.1 and 1.0 arcsec and $G$-band contrasts of 3 magnitudes or less \citep{Wood2021}. Finally, we eliminated all stars with a fractional uncertainty in radius that exceeded 20\%. 

\begin{figure}
\centering
\vspace{0.5cm} 
\includegraphics[width=1\linewidth]{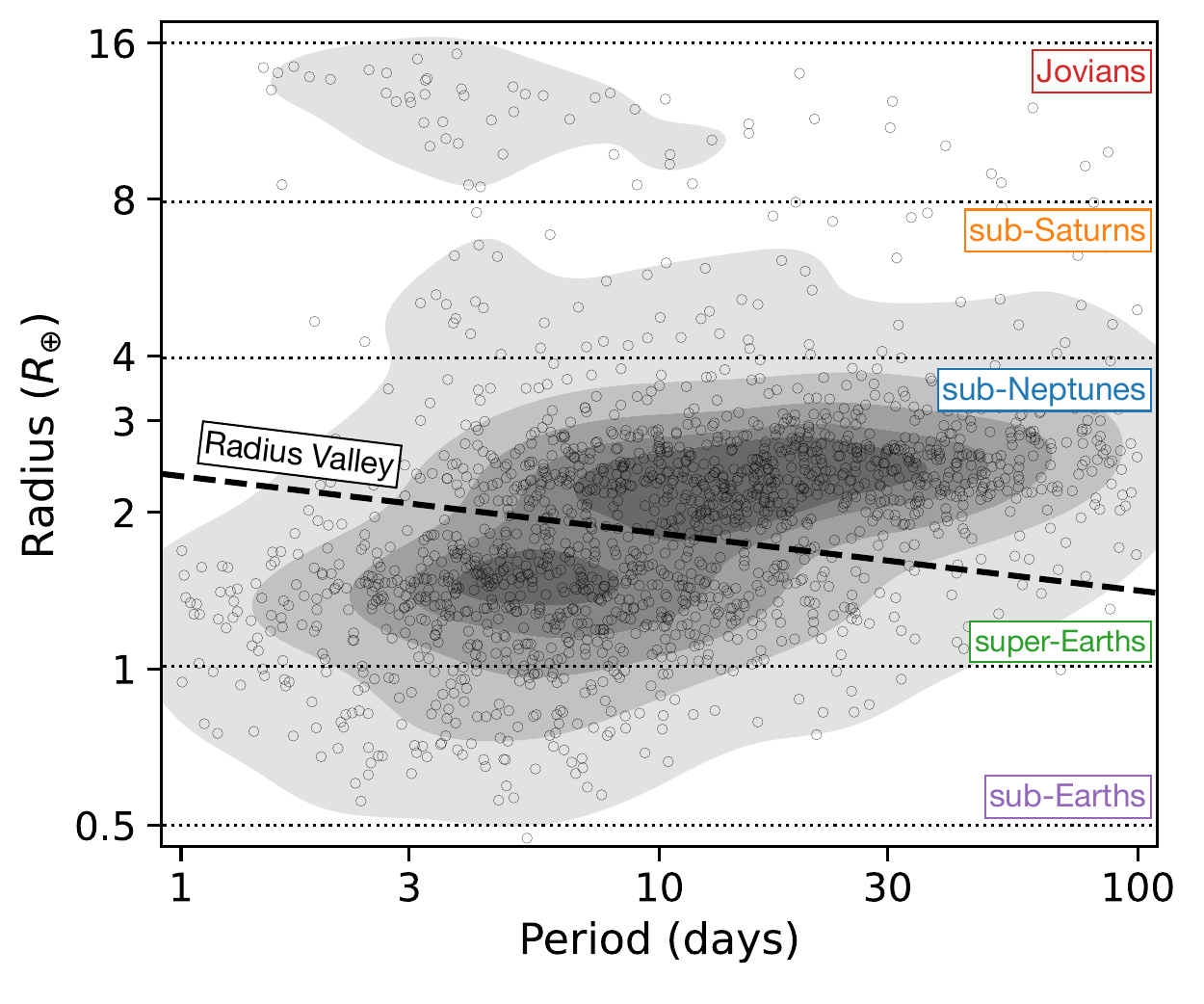}
\caption{Our sample consists of \numplanets planets with orbital periods $P = 1-100$ days and sizes $R_p = 0.5-16 R_{\oplus}$. Each circle represents a single planet, with iso-density contours drawn to guide the eye. Planets may be broadly classified based on size as sub-Earths (purple), super-Earths (green), sub-Neptunes (blue), sub-Saturns (orange), and Jovians (red). Dashed lines denote our operational definition for each group, with consistent colors adopted throughout this manuscript. The two most populous groups - super-Earths and sub-Neptunes - are separated by a ``radius valley'' whose center is given by $R_p = 1.84 (P/10\ \rm days)^{-0.11}$~\Re \citep{Petigura2022}. Planets below the radius valley are mostly rocky, whereas those above the valley possess cores with sufficient gravity to maintain massive gas envelopes.}
\label{fig:population}
\end{figure}

Throughout this work, we considered only planets between $P = 1-100$ days and $R_p = 0.5 - 16$~\Re (corresponding roughly to Mars- to Jupiter-size objects). All together, these restrictions reduced our sample to \numplanets planets and \numstars stars (Figure \ref{fig:population}). Roughly half of the planets (\numsingles) belong to single-transiting systems, while the other half (\nummultis) belong to multi-transiting systems.

\subsection*{Inferring the transit model from photometry}

We modeled the transit profiles using \alderaan, an open-source pipeline we developed for this and related projects.%
\footnote{\alderaan stands for ``Automated Lightcurve Detrending, Exoplanet Recovery, and Analysis of Autocorrelated Noise.'' The full pipeline can be accessed at \url{www.github.com/gjgilbert/alderaan}.}
\alderaan takes in raw Kepler photometric observations to produce posterior samples of period $P$, transit epoch $t_0$, planet-to-star radius ratio $R_p/R_\star$, impact parameter $b$, and first-to-fourth contact transit duration $T$. It also models transit timing variations (TTVs, i.e. deviations from a linear ephemeris) if necessary.

The first step of our procedure was to detrend (pre-whiten) the photometric lightcurves. To maintain homogeneity in our dataset, we used long cadence (30 min) photometry for all targets. We began with the pre-search data conditioning simple aperture photometry (\citealp{Stumpe2012}), accessed via the Mikulski Archive for Space Telescopes. We removed residual instrumental noise and astrophysical stellar variability using a Gaussian Process (GP) regression. \alderaan automatically detected gaps in time (i.e. missing data) or jumps in flux (i.e. discontinuities in brightness values) and treated each disjoint section of data independently. To avoid fitting out the transit profile with the GP model, we excluded three times the total transit duration from the regression. Transit durations were determined by querying the DR25 catalog \citep{Thompson2018} and TTVs (where applicable) were determined by querying a uniform TTV catalog \citep{Holczer2016}.

Following detrending, our next step was to produce a self-consistent estimate of the transit parameters $\{P, R_p/R_\star, b, T\}$ and TTVs. Accurate determination of TTVs is important because unaccounted for TTVs will smear out transit ingress and egress and bias eccentricity measurements (\citealp{Kipping2014-asterodensity, VanEylen2015}). The two issues are coupled: accurate determination of the transit shape is needed to infer accurate TTVs, and vice versa. We employed an iterative solution: first, we adopted a nominal TTV model, then we determined the transit shape, refined the TTV model, and repeated. An overview is as follows, with further details in the Supplemental Information.

We first fixed the transit times to the values in \citep{Holczer2016} and calculated the {\em maximum a posteriori} values of the transit parameters, producing a first estimate for the transit shape. Stellar limb darkening coefficients were held constant at values derived from Gaia measurements \citep{GaiaDR2, Berger2020-stars} and stellar atmosphere models \citep{Husser2013, pyldtk:2015}. We then held the transit parameters fixed and measured each transit time individually by cross-correlating the transit template across a grid of transit center time offsets $t_c$, identifying the best-fit transit time from the $\chi^2$ minimum. For low signal-to-noise transits, individual $t_c$ measurements are noisy, so we applied a regularization scheme. First, we clipped $5\sigma$ outliers, after which we tested the following suite of models: a Matern-3/2 GP, a single-component sinusoid, and polynomials of 1st, 2nd, and 3rd degree. We selected the model favored by the Akaike Information Criterion \citep{Akaike1974-AIC}.

To produce our final set of detrended lightcurves, we repeated the detrending procedure above using our self-consistent TTVs. For this second iteration, we set the width of the GP interpolation window for each known transit to $\Delta t = T + 3\sigma_{\rm TTV} + 30\ \text{min}$ where $\sigma_{\rm TTV}$ is the root-mean-squared deviation of our individually measured transit times compared to our fiducial regularized model. This interpolation window was usually much narrower than the window used during the initial detrending, which produced a better estimate of the true variability in and near transit.

To extract transit parameters, we fit each planet using a five-parameter transit model $\{c_0, c_1, R_p/R_\star, b, T\}$ where $c_0$ and $c_1$ were linear perturbations to the fiducial TTV model. Priors on all parameters were set to broad, uninformative normal ($c_0$, $c_1$), uniform ($b$) or log-uniform ($R_p/R_\star, T$) distributions. The model also included two quadratic stellar limb darkening coefficients $\{q_1, q_2\}$, following \citep{Kipping2013-limbdark} to ensure a physical profile. During sampling, we converted ${q_1,q_2}$ into the standard physical limb darkening coefficients $\{u_1,u_2\}$, where $q_1 = (u_1+u_2)^2$ and $q_2 = 0.5u_1(u_1+u_2)^{-1}$. We then applied Gaussian priors on $\{u_1, u_2\}$ with standard deviation $\sigma(u)=0.1$ and mean values informed by stellar properties; each nominal mean  was determined using Gaia stellar parameters \citep{GaiaDR2, Berger2020-stars} and PHOENIX stellar atmosphere models \citep{Husser2013, pyldtk:2015}. In multi-planet systems, we fit all planets simultaneously which involved a total of $5\times N + 2$ free parameters.

Accurately determining the credible set of transit shape parameters that are consistent with the observed data requires care. In general, the posterior probability distributions exhibit thin, curving covariances between $b$, $R_p/R_\star$, and $T$. These covariances are especially strong for grazing and near-grazing transits ($b \gtrsim 0.7$) and present a challenge for standard Monte Carlo samplers \citep{Gilbert2022-umbrella}. Previous analyses of Kepler photometry have struggled to accurately characterize the impact parameter \citep{Petigura2020, Gilbert2022-umbrella, Gilbert2022-pseudodensity}. Prior work has demonstrated that nested sampling  is well-suited to this posterior topology \citep{Skilling2004, Higson2019, dynesty:2020, MacDougall2023}.

To ensure our modeling proceeded as expected, we systematically inspected each of the \numplanets lightcurve fits. To facilitate the process, we built a custom visualization tool that displays the following data products: the phase-folded light curve, credible model draws, and model residuals; the individual transits; time series of TTVs; and posterior corner plots. We verified the detrending step by confirming that the model residuals in- and out-of-transit were white, Gaussian, and homoscedastic. We ensured the TTVs models were not over- or under-fit by direct inspection of the TTV timeseries and by checking the folded transit model for excess noise during ingress/egress (a sign of unaccounted for TTVs). We also inspected the 2D joint posteriors, and confirmed that the covariances between $b$, $T$, and $\Rp/\Rstar$ met expectations for each transits S/N and morphology (e.g., $b$ for V-shaped transits should extend beyond 1). Of the \numplanets planets, only 89 (5\%) were flagged for manual intervention. In these cases, we refit the transit model using an alternative TTV model selection routine. We found that our inferences of population eccentricity were virtually identical whether we adopted pre- or post-intervention posterior chains for these 89 planets.

\subsection*{Extracting eccentricity constraints from the transit model}

Although our model did not directly encode eccentricity, we may compare the observed duration $T$ to the predicted duration $T_0$ for a transit on a circular ($e=0$), center-crossing ($b=0$) orbit around a star with density $\rho_{\star}$ to constrain eccentricity on a planet-by-planet basis \citep{FordQuinnVeras2008, DawsonJohnson2012, Kipping2014-asterodensity}. The two durations are related via:
\begin{equation}\label{eq:photo-ecc}
    \frac{T}{T_0} \approx \sqrt{1-b^2} \left(\frac{\sqrt{1-e^2}}{1+e\sin\omega}\right).
\end{equation}
The exact expression is given in the supplemental material. Physically, the $\{e,\omega\}$ term arises from the change in orbital velocity from non-zero eccentricity, while the $\sqrt{1-b^2}$ term arises from the shortened transit chord for non-zero impact parameter.

If impact parameter is measured well, then a significant difference between $T$ and $T_0$ implies a non-zero eccentricity. For exceptionally high signal-to-noise transits this is possible because ingress/egress timescale $\tau$ encodes the impact parameter. However, for the majority of Kepler light-curves, constraints on $b$ are broad. Thus, most of the information about $b$, $e$, and $\omega$ is encoded in $T$ and the inversion is underconstrained. Nevertheless, some information can yet be extracted from even imprecise measurements of $\tau$, which is why the ``full lightcurve profiling'' method which considers joint samples $\{b, \rho_{\star}, e, \omega\}$ outperforms the simpler ``duration-only'' method.

We converted our transit-derived posterior samples of $\{P, R_p/R_{\star}, b, T\}$ to joint samples of $\{P, R_p, b, \rho_{\star}, e, \omega\}$ following the conceptual framework outlined above and implementing an importance sampling routine described in \citealp{Gilbert2022-pseudodensity} and \citealp{MacDougall2023}. A brief summary of the procedure is as follows. First, we drew random samples of $\{e, \omega\}$ from uniform distributions $e \sim \mathcal{U}(0,1)$, $\omega \sim \mathcal{U}(0,2\pi)$ and combined these with observed samples $\{P, R_p/R_{\star}, b, T\}$. For each sample, we computed physical stellar density $\rho_{\star, \rm samp}$ required to produce the observed $T$ given $\{P, R_p/R_{\star}, b, e,\omega \}$ (see \citealp{Winn2010} and Supplemental Information). We drew an identical number of stellar density $\rho_{\star}$ samples based on the Gaia constraints. Then, we computed the log-likelihood of each set of \{$\rho_{\star, \rm samp}$,$\rho_{\star}$\}:
\begin{equation}\label{eq:log-like}
    \log \mathcal{L}_i = -\frac{1}{2}\Big(\frac{\rho_{\rm \star, samp, \textit{i}} - \rho_{\star, i}}{\sigma_{\rho_{\star, i}}}\Big)^2.
\end{equation}
We converted each sample's $\mathcal{L}_i$ into an importance weight 
\begin{equation}
    w_i = \frac{\mathcal{L}_i}{\sum_i \mathcal{L}_i}.
\end{equation}
Finally, we converted $R_p/R_{\star} \rightarrow R_p$ using the Gaia-derived $R_{\star}$. To simplify downstream calculations, converted our weighted samples to unweighted samples of $\{P, R_p, b, \rho_{\star}, e, \omega\}$ for each planet.

The posterior samples generated by the above procedure are equivalent to samples which would have been generated by directly fitting an eccentric transit model to the lightcurves using Gaia-informed Gaussian priors on $\rho_{\star}$ and uninformative priors on all other parameters \citep{MacDougall2023}. The resulting posterior distributions on $e$ have a typical spread $\sigma(e){\sim}0.3$ and are asymmetric and heavy-tailed (see Supplemental Information for some examples).
 
To ensure we included the highest quality measurements, we removed 46 planets which had greater than 20\% fractional uncertainty on $R_p$. We also removed 63 planets with more than 5\% of samples met the criterion $b > 1 - R_p/R_{\star}$, as this region of parameter space contains pathological covariances which frustrate accurate sampling \citep{Gilbert2022-umbrella}. We also removed 4 planets where less than  $5\%$ of the importance weights $w_i$ exceeded machine precision; this condition demonstrates a fundamental inconsistency between Gaia stellar values and our derived transit parameters and could indicate inaccurate star and/or planet properties. We experimented with these selection criteria, finding our results were insensitive to exact cutoffs on $R_p$ precision, grazing $b$ fraction, or importance weight efficiency.

\begin{figure*}
\centering
\includegraphics[width=1\textwidth]{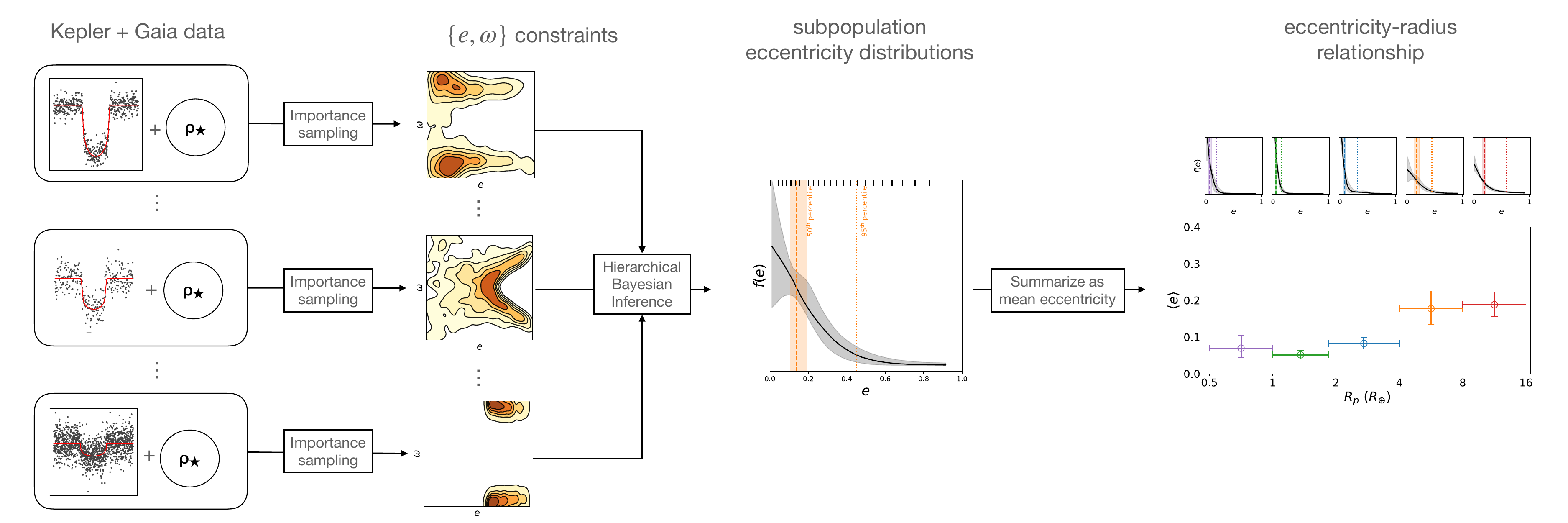}
\caption{Schematic description of our methods. (a) We began with \numplanets photometric lightcurves observed by Kepler. We fit each lightcurve using a five-parameter transit model to produce posterior samples of period $P$, transit epoch $t_0$, planet-to-star-radius ratio $R_p/R_{\star}$, impact parameter $b$, and transit duration $T$. (b) By comparing our transit-derived $T$ to the value predicted for a circular, center-crossing transit $T_0$ and applying the importance sampling scheme developed in \citealp{Gilbert2022-pseudodensity} and \citealp{MacDougall2023}, we derived joint posterior samples for eccentricity $e$ and argument of pericenter $\omega$. (c) After dividing planets into various size bins, we fit a hierarchical Bayesian model \citep{Hogg2010} to determine each sub-population's eccentricity distribution $f(e)$. We modeled the distribution non-parametrically to infer the functional form of the distribution and found a self-similar shape across planet sizes. We used this shape to construct an empirical distribution template. (d) Finally, we re-calculated the hierarchical model using our empirical template, allowing us to efficiently ``slice-and-dice'' the population by planet radius and other variables. To summarize each sub-population, we computed mean eccentricity $\e$ and its uncertainty.}
\label{fig:methods}
\end{figure*}

\subsection*{Inferring the distribution of exoplanet eccentricities}
Care is required to characterize population eccentricity distribution, given the varied and non-Gaussian nature of the eccentricity constraints of individual objects. We accomplished this using an approximate hierarchical Bayesian model, following the formalism first described by \citealp{Hogg2010}. This technique has previously been used to infer exoplanet eccentricity distributions by \citealp{VanEylen2019}, \citealp{SagearBallard2023}, and \citealp{Bowler2020}. Here, we present a brief overview of the method, with mathematical details enumerated in Supplemental Information.

Given a population of $N$ planets each with $K$ samples from its own eccentricity posterior distribution, $e_{nk}$, the combined likelihood for the population distribution $f(e)$ is
\begin{equation}\label{eq:hbayes-likelihood}
    \mathcal{L}_{\alpha} = \prod_{n=1}^N \frac{1}{K} \sum_{k=1}^K \frac{f_\alpha(e_{nk})}{p_0(e_{nk})}.
\end{equation}
Here, $p_0(e)$ is the uninformative ``interim'' prior on $e$ applied during transit modeling and $f_\alpha(e)$ is the informative ``updated'' population distribution we wish to infer; the subscript $\alpha$ denotes the vector of hyper-parameters describing $f(e)$. Conceptually, the ratio $f_\alpha/p_0$ captures the degree to which our informative eccentricity population model improves upon an uninformative prior, marginalized over all covariant quantities. While the hyperpriors on the interim distribution $p_0(e)$ are fixed, the hyperpriors on the target distribution $f_\alpha(e)$ are themselves open to inference. Determining $\{\alpha\}$ for some assumed functional form $f_\alpha(e)$ enables inference of the optimal population model.

For our specific application to eccentricity, we modified $\mathcal{L}_\alpha$ to account for geometric biases which cause preferential detection of exoplanets on eccentric orbits (\citealp{Barnes2007, Burke2008, Kipping2014-ecc-priors}). The full marginalized likelihood accounting for detection biases is
\begin{equation}\label{eq:hbayes-likelihood-with-detection-bias}
    \mathcal{L}_{\alpha} = \prod_{n=1}^N \frac{1}{K} \sum_{k=1}^K f_\alpha(e_{nk}) \left( \frac{1-e_{nk}^2}{1+e_{nk}\sin\omega_{nk}} \right)
\end{equation}
where $p_0 = 1$ because we sampled $\{e,\omega\}$ from a uniform interim prior. Omitting this geometric term would result in the measurement of transiting planet eccentricity distribution rather than the intrinsic eccentricity distribution. Prior works have adopted both conventions (see, e.g., \citealp{VanEylen2019} and \citealp{SagearBallard2023}) but the latter permits a straightforward comparison to Doppler studies. Since most eccentricities for real planets are, in fact, low, the inclusion of the detection efficiency term has only a small effect on $f_\alpha(e)$.

The true functional form of $f_\alpha(e)$ is presently unknown, although the half-normal, Rayleigh, and beta distributions are popular parametric models in the literature \citep{Kipping2013-beta, Xie2016, Mills19, VanEylen2019, SagearBallard2023}. The first two have a single free parameter, while the third has two free parameters. Imposing any of these functions at the outset runs the risk of choosing a model that cannot accurately describe the data. Therefore, we adopt a more flexible approach and use a regularized histogram instead \citep{ForemanMackey2014, Masuda2022}. The benefit of this method is that we make no assumption about the form of the distribution (unimodal, symmetric, etc.) other than that it must be smooth. We fit the distribution $f_{\alpha}(e)$ using a 25-bin histogram (resolution $\Delta e \sim 0.04$), with distribution ``smoothness'' enforced via a Gaussian process prior on the bin heights (see Figure \ref{fig:methods} for a visual representation of the histogram bin positions). In this context, our hyper-parameters $\{\alpha\}$ are the 25 logarithmic histogram bin heights, plus two parameters specifying the smoothing function.

To investigate if and how the shape of the eccentricity distribution changes as a function of planet size, we divided the planet sample into five physically motivated sub-populations in $R_p$ and derived $f_\alpha(e)$ independently for each (see Supplemental Information for details). We find that all sub-populations exhibit the same qualitative shape for $f(e)$: a mode at $e=0$ and a monotonic, quasi-exponential descent toward zero at $e=1$ (see Figure \ref{fig:methods}). Compared to the commonly used parametric distributions in the literature, our flexible distribution most resembles the beta distribution. This is consistent with the sample of Doppler measurements of giant planets, whose eccentricities are well described by a beta distribution \citep{Kipping2013-beta}. The Rayleigh distribution and half-normal distribution are poor fits at the low and high ends of the eccentricity range respectively (see Supplemental Information for discussion).

Given that the different sub-populations were described by similar $f(e)$ shapes, we adopted an empirical template for $f(e)$:
\begin{equation}\label{eq:empirical-pdf}
    \ln f_E(e;\nu,h) = \nu \left[\ln f_0(he)\right] + (1-\nu) \left[\ln f_0(0)\right].
\end{equation}
Here, $f_0(e)$ is derived by fitting a hierarchical model to our full planet sample. The empirical $f_E(e)$ has two free parameters: $\nu$ sets the tail weight, and $h$ sets the central width of the distribution. These transformations do not necessarily preserve normalization, so after transforming $f_E(e)$ we numerically re-normalized the distribution. Moving forward, we assume that $f(e)$ is self-similar across sub-populations and fit models by determining these two hyperparameters $\{\nu,h\}$.

\subsection*{Characterizing mean eccentricity for sub-populations}

Now that we have determined the shape of the eccentricity distribution, we refit various sub-populations in order to explore the covariance between $e$ and other quantities of interest, in particular $P$, $R_p$, and [Fe/H]. For each sub-population, we summarized the distribution eccentricity as $\e \pm \sigma_{\e}$, where $\e$ is a median-of-means. More specifically, for each sub-population we have generated a set of samples of $f(e)$. For each $\{\nu, h\}_i$ we generated a $f(e;\nu,h)$ and computed its mean $\e_i$. We report the median of the $\e_i$ along with their 16th and 84th percentiles to convey uncertainties. An illustration of this data compression process is shown in Figure \ref{fig:methods}.

\section*{Analysis \& Results}\label{sec:results}

Figures \ref{fig:ecc-rp}, \ref{fig:ecc-rp10-radius-gap}, and \ref{fig:period-metallicity} summarize the inter-relationships between eccentricity and planetary radius, orbital period, and stellar metallicity. We discuss each relationship below.

\subsection*{The relationship between eccentricity and planet radius}

\begin{figure}[!t]
\centering
\includegraphics[width=0.45\textwidth]{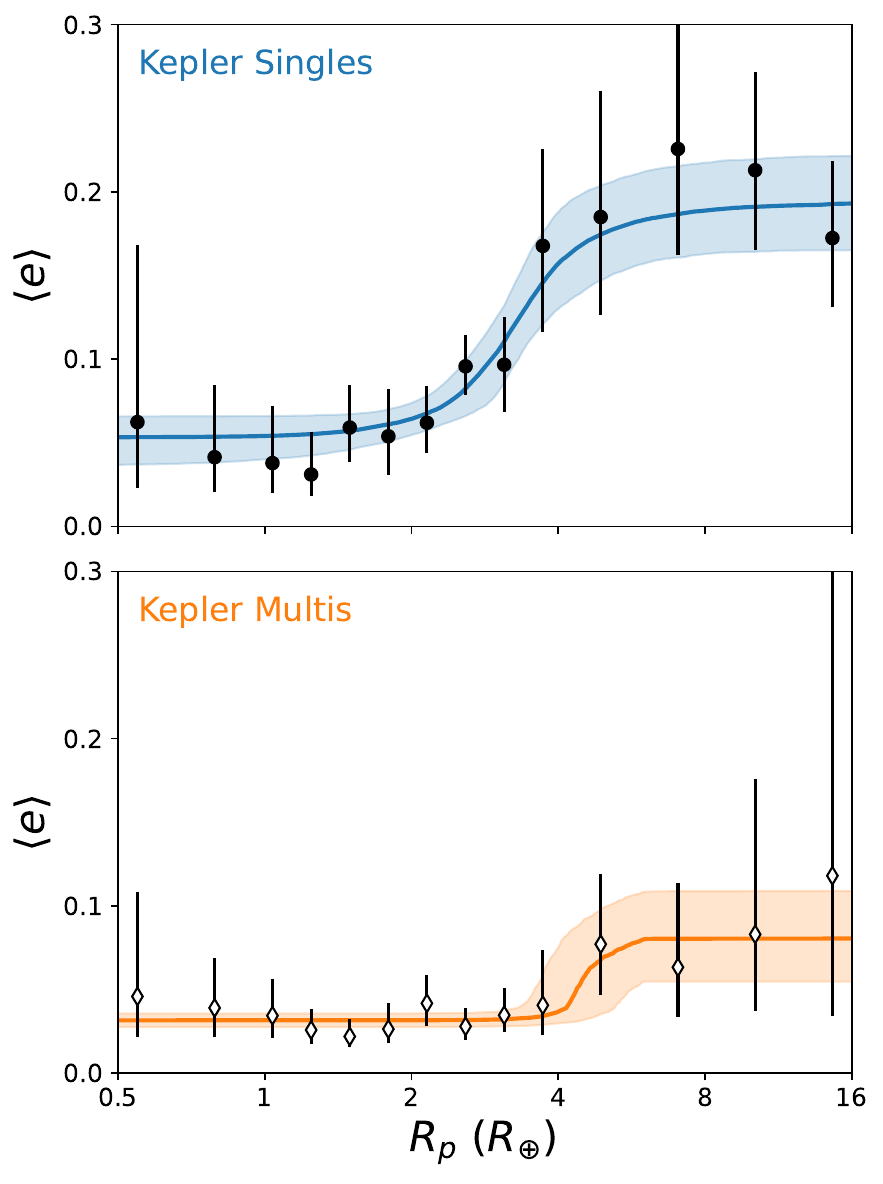}
\caption{Mean eccentricity $\e$ as a function of planet radius $R_p$ for single-transiting systems (top) and multi-transiting systems (bottom). Singles and multis exhibit a similar $\e$-$R_p$ relationship, with planets larger than Neptune having about three times the eccentricity of smaller planets. Singles are $2.4\pm0.9$ times more eccentric than multis across the full range of planet sizes.}
\label{fig:ecc-rp}
\end{figure}

To investigate the $\e-R_p$ relationship, we split the sample in the following way. First, we separated single- and multi-transiting systems (henceforth referred to as ``singles'' and ``multis''), as previous works have consistently found significant differences between these two populations \citep{Xie2016, Mills19, VanEylen2019, SagearBallard2023}. We also divided the populations into narrower bins of width in log-radius of $20\%$ for planets between $1-4 \Re$ and $30\%$ or $40\%$ for smaller/larger planets (where lower numbers motivated coarser binning, see Table \ref{tab:rp_bins}).

\begin{table}
\centering
\caption{Size bins used to determine eccentricity-radius relationship}
\begin{tabular}{rrc}
$R_0$ & $\sigma (R)$ & $f_{\rm FWHM}$\\
\midrule
$0.548$ & $0.093$ & $0.4$ \\
$0.789$ & $0.101$ & $0.3$ \\
$1.037$ & $0.088$ & $0.2$ \\
$1.245$ & $0.106$ & $0.2$ \\
$1.494$ & $0.127$ & $0.2$ \\
$1.793$ & $0.152$ & $0.2$ \\
$2.151$ & $0.183$ & $0.2$ \\
$2.582$ & $0.219$ & $0.2$ \\
$3.099$ & $0.263$ & $0.2$ \\
$3.719$ & $0.316$ & $0.2$ \\
$4.889$ & $0.622$ & $0.3$ \\
$7.041$ & $1.196$ & $0.4$ \\
$10.141$ & $1.722$ & $0.4$ \\
$14.605$ & $2.481$ & $0.4$ \\
\bottomrule
\end{tabular}

\addtabletext{To determine the $\e$-$R_p$ relationship, we divided planets into fourteen size bins, where $R_0$ are the bin centers, $\sigma(R)$ are the bin widths, and $f_{\rm FWHM}$ is the fractional full-width-half-max corresponding to $\sigma (R)/R$. See Equation \ref{eq:gaussian_R} and corresponding text for mathematical formulation.}
\label{tab:rp_bins}
\end{table}

Because most planets had fractional radius uncertainties $\sigma(R_p){\sim}10\%$ that were comparable to bin size, we followed \citep{NewtonRafferty1994} and modified the hierarchical Bayesian likelihood (Equation~\ref{eq:hbayes-likelihood-with-detection-bias}) to account for their size uncertainties:
\begin{equation}\label{eq:hbayes-likelihood-weighted}
    \mathcal{L}_{\alpha} = \prod_{n=1}^N \Bigg[ \frac{1}{K} \sum_{k=1}^K f_\alpha(e_{nk}) \left( \frac{1-e_{nk}^2}{1+e_{nk}\sin\omega_{nk}} \right) \Bigg]^{w_n}.
\end{equation}
We computed the weights $w_n$ as follows: First, we defined each bin as a Gaussian
\begin{equation}\label{eq:gaussian_R}
    g(R) = \frac{1}{\sigma_R\sqrt{2\pi}}\exp\left(-(R-R_0)^2/2\sigma_R^2\right)
\end{equation}
where $R_0$ is the bin center, and $\sigma_R$ is the bin width. We then calculated each planet's weight by summing over all $K$ samples as
\begin{equation}\label{eq:planet_weights}
    w_n = \frac{1}{K}\sum_{k=1}^K g(R_{nk}).
\end{equation}
We then normalized the weights so that the expected value $\langle w \rangle = 1$.

We found that fits using the weighted likelihood (Equation~\ref{eq:hbayes-likelihood-weighted}) were consistent with the unweighted likelihood (Equation~\ref{eq:hbayes-likelihood-with-detection-bias}). We elected to use the former because it incorporated radius uncertainties that are comparable to the narrowest bin widths.

Both singles and multis exhibit a similar relationship between $\e$ and $R_p$ (Figure \ref{fig:ecc-rp}). Eccentricity is low between $0.5-2.0$~\Re and then rises to an elevated plateau above $6$~\Re. The transition between low and high eccentricity is somewhat more gradual for singles. Overall, $\e$ is higher for singles than for multis, consistent with previous analyses \citep{Xie2016, Mills19, VanEylen2019, SagearBallard2023}.

To quantify the transition feature in this relationship, we fit the $\e$-\Rp curve with a logistic sigmoid
\begin{equation}
    f(x) = B + \frac{L}{1+e^{-k(x-x_t)}}.
\end{equation}
Here, $x\equiv\log R_p$, $B$ encodes the baseline eccentricity, $L$ encodes the overall normalization, and $\{x_t, k\}$ encodes the location and rate of change from low to high $e$. In physical terms, $e_{\rm low} = B$ and $e_{\rm high} = L + B$.

The shape of the $\e$-\Rp curve is similar for singles and multis. The transition from low-to-high $e$ occurs at $R_p = 3.3 \pm 0.4$~\Re for singles and at $R_p = 4.2 \pm 0.9$~\Re for multis. For singles, $e_{\rm high}/e_{\rm low} = 3.9^{+1.9}_{-1.0}$ while for multis $e_{\rm high}/e_{\rm low} = 2.5^{+1.0}_{-0.8}$. Thus, both the location and amplitude of the transition agree within $1\sigma$ between singles and multis.

Overall, the mean eccentricity of singles is higher than that of multis by a factor of $2.4^{+1.5}_{-0.7}$. In the flat region of the curve on the small planet end ($R_p < 1.5~$\Re), we compute $e_{\rm singles}/e_{\rm multis} = 1.7^{+0.5}_{-0.4}$, while on the large planet end ($R_p > 6.0$~\Re) we compute $e_{\rm singles}/e_{\rm multis} = 2.5^{+1.4}_{-0.7}$. 

For intermediate sizes where the curves are most visually discrepant, $e_{\rm singles}/e_{\rm multis}$ ranges from 1.3 to 6.2 ($95\%$ confidence). Thus, the single-to-multi enhancement factor for eccentricity is consistent with a constant (or nearly constant) value across the full range of planet sizes.

This result is broadly consistent with measured single-to-multi enhancement factors of $8\pm7$ \citep{Xie2016}, $4.8\pm1.8$ \citep{Mills19}$, 3.9\pm1.1$ \citep{VanEylen2019}, and $6.8\pm3.7$ \citep{SagearBallard2023}. However, previous works investigated different samples of planets, so one should not necessarily expect the offsets to be identical.

\begin{figure}[!t]
\centering
\includegraphics[width=0.45\textwidth]{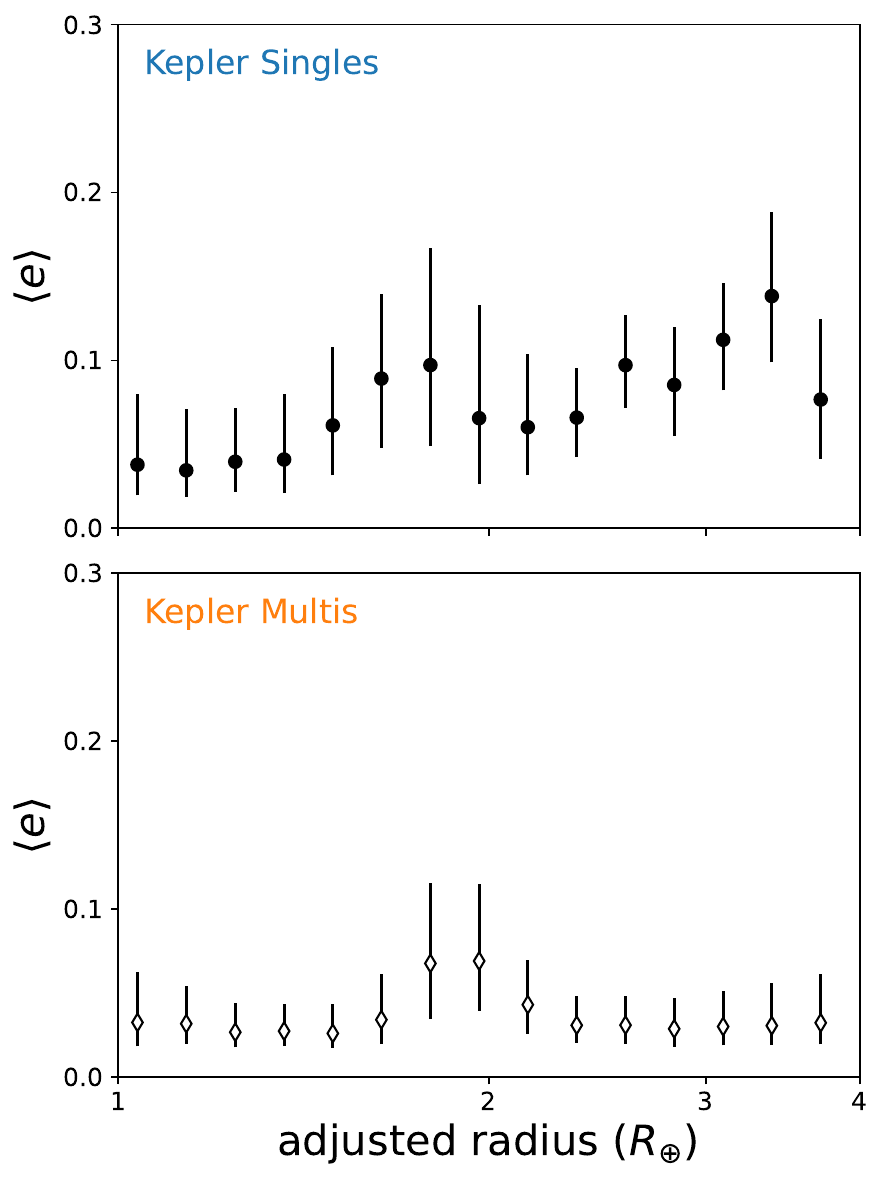}
\caption{Mean eccentricity $\e$ as a function of adjusted planet radius near the radius gap for single-transiting systems (top) and multi-transiting systems (bottom). The ``adjusted radius'' is calculated following \citealp{Petigura2022} and \citealp{Ho2023} to account for the period dependence of the radius gap. Planets in the gap show tentative evidence of elevated $\e$ compared to planets on either side of the gap.}
\label{fig:ecc-rp10-radius-gap}
\end{figure}

Small planets are bifurcated into two populations: rocky super-Earths ($R_p \lesssim 1.5~\Re$) and gas-rich sub-Neptunes ($R_p \gtrsim 2.0$~\Re), with few planets in between, a demographic feature known as the ``radius gap'' or the ``radius valley'' \citep{Fulton2017}. More precisely, the radius gap is a 2D feature in $P$-$R_p$ space (see Figure \ref{fig:population}). The center of the gap is a gently declining function of orbital period (see Figure \ref{fig:population}) and roughly follows a power-law $R_{\rm p,valley} \approx 1.8~\Re$ ($P$/10 days)$^\alpha$, where $\alpha = -0.11$ \citep{Petigura2022}. Thus, to investigate the eccentricities of planets in and around the radius gap, we separated planets into bins sliced parallel to the radius gap, following \citealp{Ho2023}. We find tentative evidence that the $\e$ for planets in the radius gap are elevated compared to planets on either side (Figure \ref{fig:ecc-rp10-radius-gap}). This feature is detected at ${\sim}2\sigma$ confidence, and the physical implications are discussed below.

One complication is that planets that appear to reside in the gap may have scattered there due to $R_p/R_\star$ measurement errors. However, as discussed, $R_p/R_\star$ is covariant with $b$ and ultimately with $e$. To explore possible confounding effects, we ran a suite of injection/recovery experiments described in the Supplemental Information. We found these effects, by themselves, are insufficient to produce the observed peak at its observed significance.

\subsection*{The relationship between eccentricity and stellar metallicity}

\begin{figure*}
\centering
\includegraphics[width=0.95\textwidth]{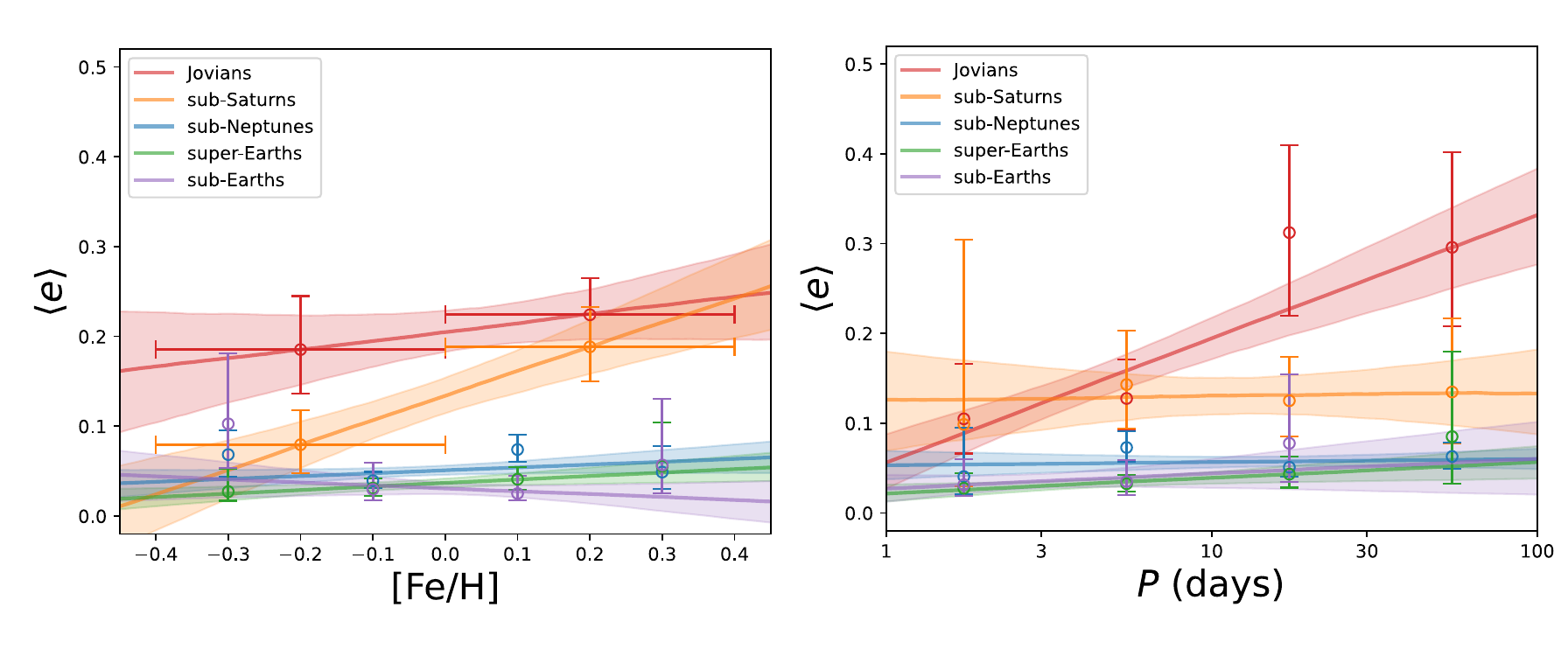}
\vspace{-0.5cm}
\caption{Mean eccentricity $\e$ as a function of (a) host star metallicity [Fe/H] and (b) orbital period $P$ for each planet size class. Planets larger than 4 Earth-radii exhibit elevated eccentricities compared to their smaller counterparts across all values of [Fe/H] and $P$. \textit{Left}: small planets exhibit a weak association between [Fe/H] and $\e$, whereas large planets exhibit a modest positive correlation. \textit{Right}: associations between eccentricity and period depend on planet size. Jovians exhibit a positive correlation between $\e$ and $P$, sub-Saturns are consistent with no correlation at a consistently elevated $\e$, and small planets are consistent with no correlation at low $\e$.}
\label{fig:period-metallicity}
\end{figure*}

A star's metallicity is its inventory of elements heavier than hydrogen and helium. It is parameterized as [Fe/H] = $\log_{10}( n_X / n_{X,\odot})$ where $n_X$ is the number density of all elements heavier than hydrogen and helium and $n_{X,\odot}$ is the solar value. Metals constitute the bulk of sub-Jovian  planets by mass and form the cores of giant planets, so it is natural to explore whether eccentricity is associated with metallicity.

Figure \ref{fig:period-metallicity} (left) shows mean eccentricity $\e$ as a function of stellar metallicity [Fe/H] for different planet size classes. For the three smallest size classes, we divided planets into four equal size bins spanning $-0.4$~dex to $+0.4$~dex. Due to the low numbers of Jovians and sub-Saturns, we divided them into two bins corresponding to sub- and super-solar metallicities.

After determining $\e$ for each of the [Fe/H]-$R_p$ bins, we fit a linear model 
\begin{equation}\label{eq:e-feh}
    \e = m \times [\text{Fe/H}] + b
\end{equation} 
to assess the strength and significance of any relationships between $\e$ and [Fe/H]. 

Eccentricity is consistently higher for large planets ($R_p > 4$~\Re) compared to small planets ($R_p < 4$~\Re) across all values of [Fe/H]. For the three small planet size classes, we do not observe any strong metallicity trends. For sub-Saturns and Jovians, however, we find a modest positive correlation between [Fe/H] and $\e$. The trend for sub-Saturns is significant at the $3\sigma$ level, while the trend for Jovians is consistent with a flat line within $1\sigma$.

\subsection*{The relationship between eccentricity and orbital period}

Figure \ref{fig:period-metallicity} (right) shows mean eccentricity $\e$ as a function of orbital period $P$ for different planet size classes. The association between $\e$ and $P$ varies depending on planet size. We quantified the strength of these $\e-P$ relationships by fitting the following model
\begin{equation}\label{eq:e-P}
    \e = m \times \log P + b.
\end{equation} 
For the largest planets (Jovians), we find a positive correlation between $\e$ and $P$, with planets inside $P<3$ days having $\e\sim0.1$, rising to $\e\sim0.3$ for planets beyond $P>10$ days. For sub-Saturns, the trend is consistent with a constant value $\e \sim 0.1$ across periods. For small planets ($R_p < 4$~\Re) we find no apparent correlation between period and eccentricity.

\section*{Discussion}\label{sec:discussion}

We detected numerous statistically significant features in the relationships between $\e$, $R_p$, $P$, and [Fe/H]. We discuss the astrophysical implications of these relationships below.

\subsection*{Distinct formation pathways for planets larger or smaller than 4 Earth-radii}

\begin{SCfigure*}
\centering
\includegraphics[width=0.65\textwidth]{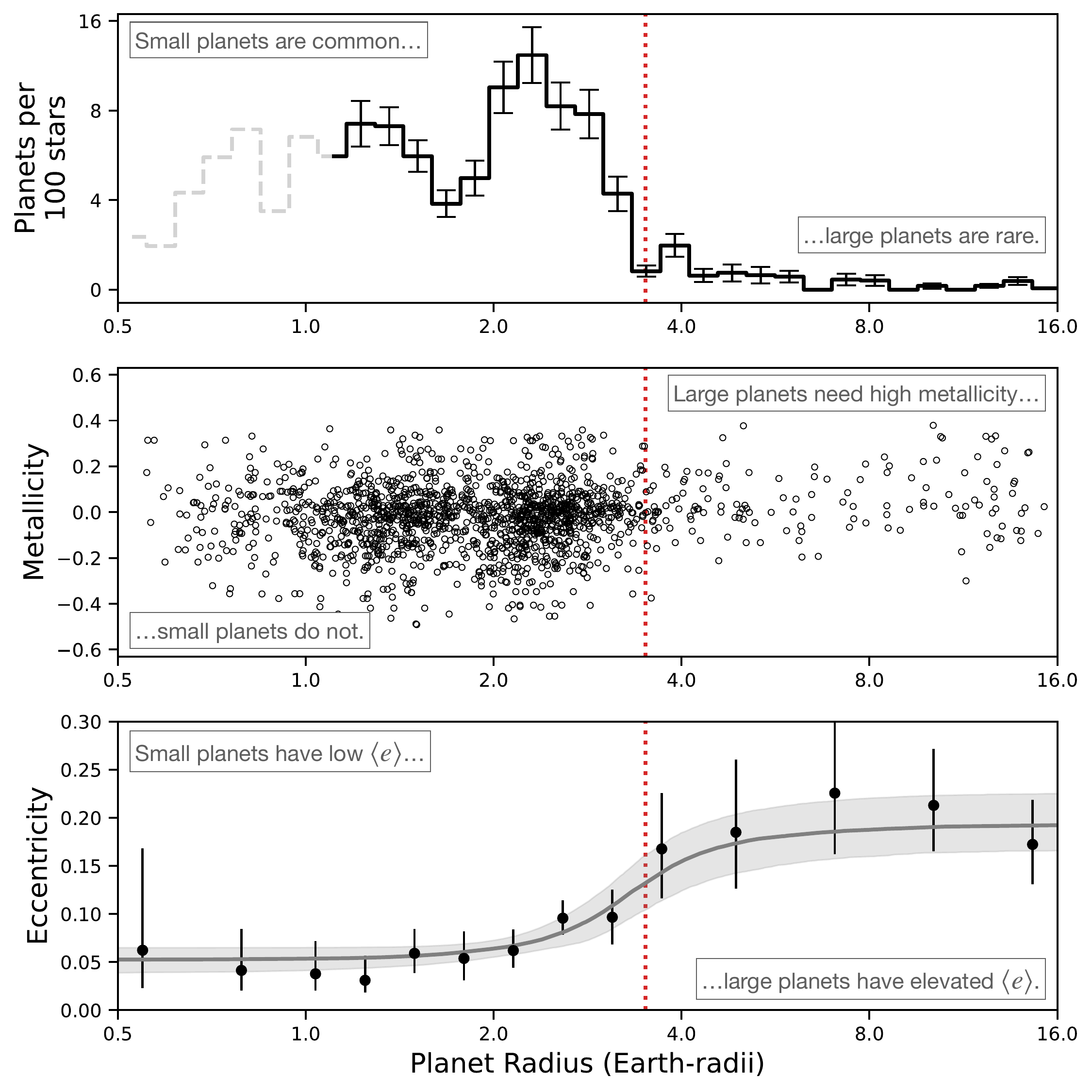}
\caption{Planets above and below a threshold of ${\sim}3.5$ Earth-radii possess qualitatively distinct occurrence rates, host star metallicities, and eccentricities. \textit{Top}: small planets are common, but larger planets are rare. Data reproduced with permission from \citealp{FultonPetigura2018}. \textit{Middle}: small planets can be found orbiting stars of any metallicity, but large planets prefer high metallicity hosts \citep[see][]{Buchhave2012}. \textit{Bottom}: Small planets tend to have low mean eccentricities $\e{\approx}0.05$, whereas large planets have elevated mean eccentricities $\e{\approx}0.2$. Taken together, these trends suggest distinct formation pathways for planets smaller/larger than ${\sim}3.5$ Earth-radii.}
\label{fig:3RE-transition}
\end{SCfigure*}

There is a sharp break in the eccentricity distribution of small vs large planets.  Intriguingly, the transition radius is consistent with several other features in the demographics of close-in planets. There is a precipitous drop in exoplanet occurrence rates above ${\sim}3.5$~\Re \citep{FultonPetigura2018}. Furthermore, planets above ${\sim}4$~\Re are strongly associated with high stellar metallicity, while planets below this threshold are not \citep{Buchhave2012,Petigura2018}. Stated plainly, large planets are rare, but those that do exist tend to be found with elevated eccentricities $\e \approx 0.2$ around metal-rich stars ([Fe/H] $> 0$~dex).  In contrast, small planets are common, possess low eccentricities $\e \approx 0.05$ and show no apparent preference for high or low metallicity hosts (Figure~\ref{fig:3RE-transition}).

The conspicuous alignment of a sharp transition in planet occurrence, eccentricity, and host star metallicity at the same planetary size threshold suggests that large planets have experienced a distinct formation pathway compared to small planets. From inspection of occurrence rate alone (Figure~\ref{fig:3RE-transition}, top), it is clear that small planets form more readily than large planets. While these small planets form across a wide range of metallicities between [Fe/H] = $-0.5$ to $+0.5$, increasing metallicity does not seem to boost the number of small planets \citep{Petigura2018}. In contrast, large planets are associated with high metallicity, but having high metallicity does not guarantee that a system will form a large planet. Indeed, \citealp{FischerValenti2005} found that only 10\% of Sun-like stars with [Fe/H] = 0.25 dex have a giant planet companion inside $P \leq 4$ years. This incongruity may be understood by considering the stochastic nature of planet formation.

Planets with cores that are between 1 and 10~$\Me$ and orbital periods less than a year occur at a rate of one per star \citep{Petigura2018}. The fraction of stars with such cores is more uncertain but estimates range from 30 to 50\% \citep{Petigura2013, Zhu2018, Hsu2019}. However, only a small fraction of these cores are able accrete sufficient H/He envelopes to become super-Neptune-sized planets (with envelope mass fractions of $f_{\rm env} \gtrsim 10\%$) before the circumstellar gas disk disperses ($t_{\rm disk} \approx 1$--10 Myr; see, e.g., \citealp{WilliamsCieza2011}). Increasing metallicity clearly increases this small fraction. However, making a large planet may require a confluence of other factors besides high metallicity such as a high accretion rate, long disk lifetime, high local gas disk mass. We do not expect a critical [Fe/H] that guarantees the formation of a large planet. 

How then does eccentricity fit into this picture? Small planets almost universally have low eccentricities $e\lesssim0.1$. These planets form within the circumstellar disk before the gas disperses, so damping will rapidly circularize any initial non-zero eccentricities. In contrast, some giant planets have elevated eccentricities $e\gtrsim0.3$; these may result from self-excitation through secular interactions and/or from planet-planet scattering events. Other giant planets have low eccentricities $e\sim0$; perhaps they form with low $e$ and planet-planet interactions are insufficient to excite them, or perhaps they are excited while the gas disk is sill present and there is sufficient damping to circularize. The moderate mean eccentricity observed among large planets would then be the average of low-$e$ and high-$e$ formation histories.

Because small planets form at both high and low metallicities, we should expect there to be some systems of small planets in which eccentric giant planet formation also occurred. In this case, there is a possibility that the eccentric giants would excite their smaller siblings. However, occurrence rates suggest that these outcomes are rare. Among the high metallicity stars, only 10\% produce giant planets; of those, only a fraction are high-$e$; and, of those, only a fraction have orbits that can effectively transfer eccentricity to the inner system. We estimate that the fraction of small planets with distant-giant-excited eccentricities to be no more than about 5\%. Even if every small planet in such systems were highly eccentric ($e\gtrsim0.3$), the preponderance of low-$e$ small planets means that any observed $\e$ integrated over the population will also be low. This ``eccentricity dilution'' is strong enough that $\e$ for small planets may be kept low even around high [Fe/H] stars. Conversely, if high-$e$ small planets are identified, this might indicate the presence of a (perhaps unseen) giant outer companion.

\subsection*{An eccentricity peak in the radius valley}

Planets in the radius valley exhibit somewhat elevated mean eccentricity compared to planets on either side of the valley. Although the effect only marginally significant ($2.1\sigma$, see Supplemental Information for calculation), several lines of evidence suggest that the peak may indeed be physical in origin, rather than a spurious fluctuation arising from the small number of planets in the radius valley.

Numerous previous analyses have endeavored to ``clear out'' the radius valley by improving measurements of stellar and planetary radii \citep[e.g.][]{Petigura2020, Petigura2022, Ho2023}, but none has produced an entirely pristine radius valley, implying that an astrophysical population of ``gap planets'' does indeed exist. Proposed mechanisms for producing such planets include variable mass-loss rates due to the intrinsic scatter in the XUV output of host stars, a diversity of core composition (i.e. different admixtures of iron, rock, and water), or planet-planet collisions \cite{Ho2023}. Furthermore, recent efforts have detected anomalous patterns in the size ratios \citep{ChanceBallard2024} and resonance behaviors \citep{ChanceBallard2024, Dai2024} of purported gap planets, implying they may have experienced unusual dynamical histories. Any processes which would have disrupted the usual size and spacing regularity found in super-Earth and sub-Neptune systems would also be likely to have excited eccentricities. While no single line of evidence conclusively demonstrates the existence of an exotic population of gap planets, the existing measurements can be readily synthesized into a coherent picture.

If the eccentricity peak in the radius valley is indeed real, we hypothesize that the population could be generated by planet-planet collisions in post-accretion systems. Such objects could be built either from the bottom-up via mergers or from the top-down via atmospheric stripping. In the first scenario, consider two 5~\Me super-Earths with an Earth-like bulk composition. Such objects would have a radius of $\Rp = 1.0~\Re (\Mp/\Me)^{0.25} = 1.5~\Re$ \citep{OwenWu2017}, and the merger of these two objects would result in a 1.8~\Re planet. In the second scenario, consider a sub-Neptune with a $10 \Me$ Earth composition core and a $3\%$ by mass H/He envelope. This object would have $R_p \approx 3.6$~\Re; the large change in $R_p$ occurs because a small fractional mass of H/He can greatly increase a planet's radius \citep[e.g.,][]{LopezFortney2014}. An impacting body with mass $M = 0.1 \Me$ would have sufficient kinetic energy to unbind the atmosphere, leaving a bare rock with $R_p = 1.8$~\Re. In either scenario, at $P=20$ days around a Sun-like star, the final planet would fall squarely in the middle of the radius valley and would possess a non-existent or tenuous atmospheres and elevated eccentricity. These predictions can be tested with a focused Doppler campaign using existing extreme precision radial velocity instruments or with secondary eclipse measurements using JWST.

\subsection*{A common parent population for single- and multi-transiting systems}

Early analyses of the Kepler census found an overabundance of single-transiting systems compared to models developed from the statistics of multi-transiting systems \citep{Lissauer2011}. The origin of this so-called ``Kepler dichotomy'' has remained a mystery, with both one- and two- population models being invoked to explain the data \citep[e.g.,][]{Lissauer2011, Ballard2016, Zhu2018, He2020, Millholland2021}. While it is possible that an intrinsic population of singles exists, the dichotomy may instead result from observational biases. One plausible explanation is that the single-transiting systems belong to a dynamically hot population with elevated eccentricities and mutual inclinations, which produces fewer observations of multi-transiting systems even if intrinsic multiplicity is high \citep[e.g.,][]{Lissauer2011}. Alternatively, \citealp{Zink2019} explained the Kepler dichotomy as a byproduct of suppressed transit detection efficiency in multi-planet systems.

The similarity of the $\e$-$R_p$ curves for singles and multis (Figure \ref{fig:ecc-rp}) points to a common parent population and formation channel. If the single transiting systems represent some distinct population of planets with a unique formation history, we would expect a different shape to the $\e$-$R_p$ relationship. We therefore interpret Kepler's single-transiting systems as the high-$e$ high-$\Delta i$ tail of the intrinsic multiplanet population, rather than a qualitatively distinct group. 

$N$-body models of planetesimal growth have found a tendency toward equipartition of random orbital velocities which produces a tight correlation between eccentricity and inclination $\e \simeq 2 \Delta i$ \citep{Kokubo2005}. This tendency has been observed over a wide range of size and angular momentum scales that includes the Kepler population \citep{Xie2016}. Thus, eccentricity (which has a weak effect on observed multiplicity) can be used as a proxy for mutual inclination (which has a strong effect on observed multiplicity). Our observed eccentricity enhancement factor $e_{\rm singles}/e_{\rm multis} \sim 2.5$ therefore implies a similar disparity in mutual inclinations which biases the number of detected planets in dynamically hot vs dynamically cool systems.

\section*{Conclusions}\label{sec:conclusions}
In this work, we measured the distribution of eccentricities for close-in planets based on \numplanets individual planets discovered by Kepler. We summarize our main findings below:

\begin{enumerate}
    \item The eccentricity distribution $f(e)$ peaks at $e=0$ and falls toward zero at $e=1$. The shape --- but not the spread --- is consistent across a range of planet sizes. Of the commonly used parametric models in the literature, the beta distribution provides a good approximation, but the Rayleigh and half-normal distributions do not.

    \item On average, large planets are $\sim$3--4$\times$ more eccentric than small planets. The transition between low-to-high $\e$ at $R_p \sim 3.5$~\Re matches other features in the exoplanet population, specifically transitions in planet occurrence rate and host star metallicity. Taken together, these trends suggest that large and small planets form via two distinct formation channels.

    \item Planets in the the radius valley show tentative evidence of having elevated eccentricities compared to slightly larger or smaller planets. Such planets may have experienced major planet-planet collisions in their past which substantially altered their core mass and envelope fraction, moving them into the sparsely populated radius valley.

    \item  Planets in single- vs multi-transiting systems have the same $\e$-$R_p$ relation, except that, on average, singles are $\sim$2.5$\times$ more eccentric than multis. This correspondence suggests that single-transiting and multi-transiting systems belong to the same parent population and experienced similar formation histories.

    \item Eccentricity and metallicity lack a strong correlation (positive or negative). However, subtle correlations among sub-populations may exist and warrant further investigation.

\end{enumerate}

Planet eccentricities are related to other planetary properties in intricate, subtle, and surprising ways. Surely, fundamental aspects of planet formation physics are encoded in these relationships. We eagerly await new models that can account for these patterns. 

\acknow{We acknowledge the heroic efforts by the Kepler and Gaia teams, which provided the starting datasets for this project. We benefited from insightful discussions with Sarah Ballard, Konstantin Batygin, Brendan Bowler, Dan Foreman-Mackey, Mason MacDougall, James Rogers, Jason Rowe, Sheila Sagear, Hilke Schlichting, Vincent Van Eylen, Josh Winn, and Jon Zink. Funding for this work was provided by a UCLA set-up award to E.A.P. and by the Heising-Simons Foundation award \#2022-3833. We thank the anonymous referees for suggestions which greatly improved the quality of the analysis.} 

\showacknow{} 



\section*{References}\label{sec:references}
\bibliography{eccentricity}

\newpage
\onecolumn

\begin{center}
\includegraphics[width=9.95cm]{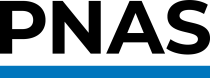}
\end{center}

\vskip45pt
\begingroup
\raggedright
{\Huge\sffamily\bfseries Supporting Information for\par}
\bigskip
{\LARGE\sffamily\bfseries Planets larger than Neptune have Elevated Eccentricities \par}
\bigskip
{\sffamily\bfseries Gregory J. Gilbert, Erik A. Petigura, and Paige M. Entrican \par\bigskip Gregory J. Gilbert. \par Email: gjgilbert@astro.ucla.edu \par}
\endgroup
\bigskip
\section*{This PDF file includes:}
\begin{list}{}{%
\setlength\leftmargin{2em}%
\setlength\itemsep{0pt}%
\setlength\parsep{0pt}}
\item Supporting text
\item Figs.~S1 to S14
\item Tables S1 to S3
\item SI References
\end{list}

\newpage
\onecolumn

\section*{Modeling transit timing variations}

Accurate determination of transit timing variations (TTVs) is important because unaccounted for TTVs will smear out transit ingress and egress and bias eccentricity measurements (\citealp{Kipping2014-asterodensity, VanEylen2015}). To measure TTVs, we adopted an iterative template-matching approach similar to the one used by \citep{Mazeh2013} and \citep{Holczer2016}. The basic idea behind this method is to first estimate the transit parameters $\{P, R_p/R_\star, b, T\}$ using a fixed ephemeris, after which the transit parameters are fixed and the individual mid-transit times $t_c$ are varied for the individual transits. Even if there is considerable uncertainty on transit shape parameters, the transit shape itself is often narrowly constrained, so adopting fixed transit parameters does not strongly bias inference of $t_c$.

We first fixed the transit times to the values in \citep{Holczer2016} and calculated the {\em maximum a posteriori} values of the transit parameters, producing a first estimate for the transit shape. We held stellar limb-darkening coefficients were at values derived from Gaia measurements \citep{GaiaDR2, Berger2020-stars} and stellar atmosphere models \citep{Husser2013, pyldtk:2015}. We then held the transit parameters fixed and measured each transit time individually by cross-correlating the transit template across a grid of transit center time offsets $\Delta t_c$, identifying the best-fit transit time from the $\chi^2$ minimum. For low signal-to-noise transits, individual $t_c$ measurements are noisy, so we applied a regularization scheme to avoid over-fitting. First, we clipped $5\sigma$ outliers, after which we tested the following suite of models: a Matern-3/2 GP, a single-component sinusoid, and polynomials of 1st, 2nd, and 3rd degree. We selected the model favored by the Akaike Information Criterion \citep{Akaike1974-AIC}. This regularization routine produced our fiducial transit timing model. Deviations from a linear ephemeris were typically on the order of minutes, although in a few cases TTV amplitudes varied by several hours.

We then refined the transit shape parameters. To extract transit parameters, we fit each planet using a five-parameter transit model $\{c_0, c_1, R_p/R_\star, b, T\}$ where $c_0$ and $c_1$ were linear perturbations to the fiducial transit timing model. More specifically, we first scale individual transit times $t_i$ to the range $(-1,1)$ such that
\begin{equation}\label{eq:legendre_x}
    x_i =\frac{2(t_i - t_{\rm min})}{t_{\rm max} - t_{\min}} - 1.
\end{equation}
During transit modeling, we express the transit ephemeris as
\begin{equation}\label{eq:ephemeris_perturbation}
    x_i' = c_0 + c_1 x_i
\end{equation}
such that the quantities $\{c_0,c_1\}$ provide a linear perturbation to the fiducial transit model. We experimented with using higher-order polynomials to model the ephemeris, finding virtually no difference in posterior transit parameter measurements but considerably higher computation time. Although linear perturbations do not fully capture the covariance between $t_c$ and other transit parameters, we determined that our linear perturbation model offered a nearly optimal trade off between flexibility and efficiency. Our uncertainties on transit parameters are thus at least slightly underestimated, but we verified with a suite of extensive injection-and-recovery tests (see below) that these underestimates are small in magnitude and do not affect our interpretation of the results.

We visually inspected the TTV waveforms and individual fits to transits to ensure they were neither over- nor under-fit. Our automated model selection routine did a good job for nearly all the fits. However a handful required manual intervention. We found that the main results were unaffected whether or not we included these manually adjusted planets in the eccentricity analysis.

\section*{Extracting eccentricity constraints from the transit model}

To extract eccentricity information from transit-derived posterior samples $\{P, R_p/R_{\star}, b, T_{14}\}$, we followed the procedure developed in \citealp{MacDougall2023}. A summary of the procedure is described in the main text, with additional details below.

For a transiting planet, the first-to-fourth contact transit duration is
\begin{equation}\label{eq:T14-winn-ecc}
T_{\rm 14} = \frac{P}{\pi}\sin^{-1}\left(\sqrt{\frac{\left(1+R_p/R_{\star}\right)^2-b^2}{\left(\frac{G P^2 \rho_{\star}}{3 \pi}\right)^{2/3}-b^2}}\right)\frac{\sqrt{1-e^2}}{1+e\sin{\omega}}.
\end{equation}
Most planets are much smaller than their host stars and orbit at planet-star separation many times larger than the stellar radius, i.e. $R_p/R_\star \ll 1$ and $a/R_\star \gg 1$. Making these approximations, we recover the formula presented in the main text:
\begin{equation}\label{eq:photo-ecc-SI}
    \frac{T_{14}}{T_0} \approx \sqrt{1-b^2} \left(\frac{\sqrt{1-e^2}}{1+e\sin\omega}\right)
\end{equation}
where $T_0$ is the predicted transit duration for a circular ($e=0$), center-crossing ($b=0$) orbit.

We used importance sampling \citep{KloekVanDijk1978} to determine the each planet's eccentricity conditioned on its lightcurve and mean stellar density. We sampled $\{e,\omega\}$ from uniform distributions and assigned them to each set of $\{P, R_p/R_{\star}, b, T_{14}\}$. We then inverted Equation \ref{eq:T14-winn-ecc} to yield
\begin{equation}
\label{eq:rhostar-winn}
\rho_{\rm \star, samp} = \frac{3 \pi}{G P^2}\left(\frac{\left(1+R_p/R_{\star}\right)^2-b^2}{\sin^2{\left(\frac{T_{\rm 14} \pi}{P}\frac{1+e\sin{\omega}}{\sqrt{1-e^2}}\right)}}+b^2\right)^{3/2},
\end{equation}
where $\rho_{\star, \rm samp}$ is the mean stellar density implied by the $\{P, R_p/R_{\star}, b, T_{14}\}$ sampled from the lightcurve fits and the $\{e,\omega\}$ assigned from our post-hoc uniform sampling. Naturally, some of these densities will be far from the true stellar density and correspond to combinations of $\{e,\omega\}$ that are inconsistent with the measurements. Other values of $\rho_{\star, \rm samp}$ will be close to the true stellar density, indicating a $\{e,\omega\}$ pair that is consistent with the measurements. 

We quantified the agreement of the each sampled density $\rho_{\star, \rm samp}$ and the true density via the following likelihood
\begin{equation}\label{eq:importance-likelihood}
    \log \mathcal{L}_i = -\frac{1}{2}\Big(\frac{\rho_{\rm \star, samp, \textit{i}} - \rho_{\star}}{\sigma_{\rho_{\star}}}\Big)^2
\end{equation}
where $\rho_{\star}$ is the Gaia-derived stellar density \citep{Berger2020-stars}. Each set of samples was assigned an importance weight 
\begin{equation}
    w_i = \frac{\mathcal{L}_i}{\sum_i \mathcal{L}_i}.
\end{equation}
Finally, we resampled our joint posteriors $\{P, R_p, b, \rho_{\star}, e, \omega\}$ to produce a set of unweighted samples for each planet. As shown in \cite{MacDougall2023}, these samples are equivalent to those which would have been generated by directly fitting an eccentric transit model to the Kepler lightcurves using Gaia-informed Gaussian priors on $\rho_{\star}$ and uninformative (uniform or log-uniform) priors on all other parameters.

\begin{figure*}[p!]
\centering
\includegraphics[width=0.95\textwidth]{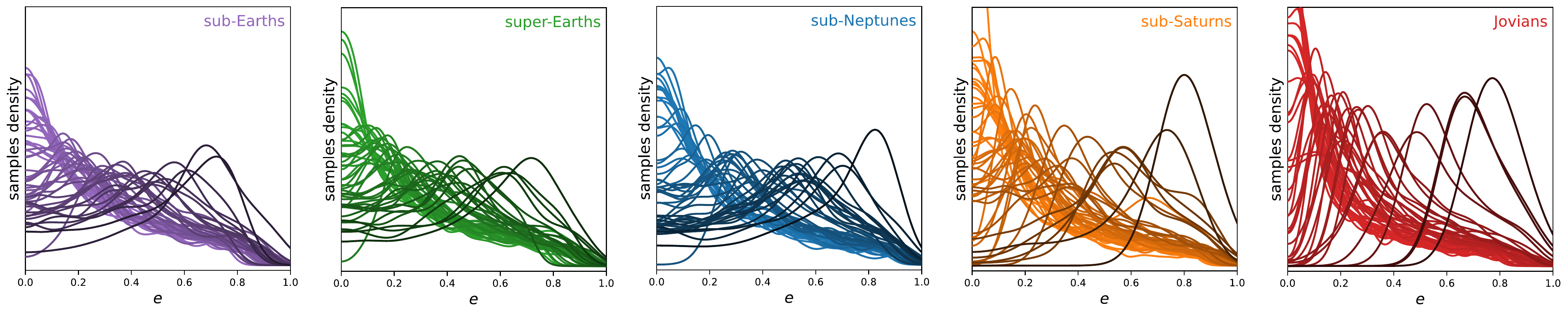}
\caption{Constraints on the eccentricities of individual planets grouped according to size class (indicated at top right). Samples of $\{e,\omega\}$ were generated via importance sampling from Kepler data. For this visualization, we randomly selected 50 planets from each planet class and calculated a Gaussian kernel density estimate of $e$. The kernel bandwidth was arbitrarily set to 0.1. By eye, it is clear that for the sub-Saturns and Jovians, there is a higher fraction of planets with confidently measured moderate or high eccentricities. More subtle differences appear when the full sample is fed into our model.}
\label{fig:ecc-posterior-kde}
\end{figure*}

\section*{Inferring the distribution of exoplanet eccentricities}

We infered $f(e)$ following the approximate hierarchical Bayesian formalism of \citealp{Hogg2010}. For specific details of modeling exoplanet eccentricity using this method we follow \citealp{VanEylen2019} and \citealp{SagearBallard2023}; for details of implementing a non-parametric likelihood we follow \citealp{ForemanMackey2014} and \citealp{Masuda2022}. A broad overview of hierarchical Bayesian analysis may be found in \citealp{Gelman13} and \citealp{Betancourt2017}. Our synthesis of these methods is as follows.

\subsection*{Hierarchical Bayesian framework}
Given a population of $N$ transit lightcurves, each of which has undergone model fitting to produce sets of $K$ eccentricity samples $e_{nk}$ \citep[see][]{Rowe2014, Rowe2015, Thompson2018, MacDougall2023}, the combined likelihood for the population distribution $f(e)$ is
\begin{equation}
    \mathcal{L}(e;\alpha) = \prod_{n=1}^N \frac{1}{K} \sum_{k=1}^K \frac{f(e_{nk}; \alpha)}{p_0(e_{nk})},
\end{equation}
where $p_0(e)$ is the uninformative ``interim'' prior on $e$ applied during model fitting (or in our case, during importance sampling) and $f(e;\alpha)$ is the informative ``updated'' population distribution we wish to infer \citep{Hogg2010}. For our purposes, $p_0(e) = 1$, since $e$ was sampled assuming a uniform prior. 

For our specific application to eccentricity, we modified $\mathcal{L}(e;\alpha)$ to account for geometric biases which favor detection of exoplanets on eccentric orbits \citep{Barnes2007, Burke2008, Kipping2014-ecc-priors}. The transit probability is proportional to
\begin{equation}\label{eq:det-prob}
    p(\hat{t}|e,\omega) = \left(\frac{1+e\sin\omega}{1-e^2}\right)
\end{equation}
where $\hat{t}$ is a boolean indicating if a planet transits. Had we not included this effect, we would be measuring the distribution of \textit{transiting} planets; by including it we measure the \textit{intrinsic} underlying distribution which may be compared to Doppler studies and other non-transit populations. The full marginalized likelihood accounting for detection biases is now
\begin{equation}
    \mathcal{L}(e; \alpha) = \prod_{n=1}^N \frac{1}{K} \sum_{k=1}^K f_\alpha(e_{nk}) \left( \frac{1-e_{nk}^2}{1+e_{nk}\sin\omega_{nk}} \right).
\end{equation}
In practice, we found that because most eccentricities for real planets are low, the inclusion of the detection efficiency term had only a small effect on our inferred $f(e)$.

\subsection*{A flexible non-parametric model}
For our specific $f_{\alpha}(e)$, we adopted a regularized histogram model. In our histogram model, the hyper-parameters $\alpha_m$ are the logarithmic bin heights in each of $M$ bins of width $\Delta e_m$, where

\begin{equation}
    \ln f(e;a,b) = 
    \begin{cases}
        \frac{1}{a-b},& \text{if } a \leq x < b \\
        0,              & \text{otherwise}.
    \end{cases}
\end{equation}

To ensure proper normalization of the probability density function we transformed $\alpha_m$ via
\begin{equation}
    \alpha_m' = \alpha_m - \ln\left({\sum_{m=1}^M \exp(\alpha_m)\Delta e_m}\right)
\end{equation}
so that $\sum_m{\exp(\alpha'_m)\Delta e_m} = 1$.

We enforced smoothness in the histogram by applying a Gaussian Process (GP) prior on $\alpha$ using a Mat\'ern-3/2 kernel. The covariance function of this kernel is
\begin{equation}\label{eq:matern32}
    \kappa(e) = s^2\left(1 + \frac{\sqrt{3}e}{\ell}\right)\exp\left(-\frac{\sqrt{3}e}{\ell}\right)
\end{equation}
where $s$ sets the scale and $\ell$ sets the correlation length. In practice, we achieved good results by applying a modestly informative log-normal hyper-priors on the GP hyperparameters $\ln(s)\sim \mathcal{N}(3,1)$ and $\ln(\ell)\sim \mathcal{N}(0,1)$. These hyper-priors prevented the sampler from wandering too far from the maximum likelihood region of parameter space and asserted a certain degree of smoothness in the distribution. Latent bin heights $\alpha$ were drawn from a standard normal distribution and scaled according to the GP prior. We experimented to determine the optimal number of eccentricity bins and found that $M=25$ bins provided a good trade-off between flexibility, robustness, and computational speed. We set the bins widths dynamically so that the $N \times K$ eccentricity samples were evenly divided among the bins. The locations of the bins are shown as ticks on the sub-population eccentricity distribution in Figure 2 in the main text. Compared to using equal bin widths, using variable bin widths provided greater resolution in the high probability density regions of the curve. A typical width for the lowest eccentricity bin was $\Delta e\sim0.02$, which sets a lower limit on measurable $\e\sim0.01$. Note, however, that after converting the regularized histogram into an empirical template (see below), subsequent transformations may produce arbitrarily small $\e$.

We implemented our model in Python using \texttt{PyMC3} \citep{pymc3:2016} and \texttt{celerite} \citep{celerite:2017}, with sampling performed using the NUTS algorithm \citep{Hoffman2011-NUTS} for gradient-based Hamiltonian Monte Carlo \citep{Neal2011-HMC}. For each model run, we drew 1000 $e_{nk}$ samples for each planet and sampled the posterior $\{\alpha, \ell, s\}$ in two chains for 1000 steps each. The target acceptance fraction was set to 0.95 and the posterior chains were inspected for convergence by requiring a Gelman-Rubin statistic $\hat{R} < 1.05$ \citep{GelmanRubin1992}.

\subsection*{An empirical estimate of the parent distribution}
The benefit of our non-parametric model is that it is flexible and does not \textit{a priori} impose a functional shape on the distribution -- the only assumption is that the distribution must be smooth. The downsides are that it is slow to compute and poorly constrained when the sample of planets is small. So, rather than applying the non-parametric model directly to sub-populations, we leveraged its flexibility to create an empirical ``template'' probability density function (PDF) using the full \numplanets planet sample. We then transformed this empirical PDF to match sub-population distributions. Our procedure is as follows.

We split the planet sample into five physically motivated bins in planet size: sub-Earths (0.5--1.0~\Re), super-Earths ($1.0~\Re$--$R_{\rm gap}$), sub-Neptunes ($R_{\rm gap}$--4~\Re), sub-Saturns (4--8~\Re), and Jovians (8--16~\Re). The center of the radius gap is given by $R_{\rm gap} = \Re (P/10\ \text{days})^{-0.11}$ following \citealp{Petigura2022}. We then independently fit our non-parametric model for $f(e)$ for each radius bin. 

\begin{figure}[p!]
    \centering
    \includegraphics[width=0.45\textwidth]{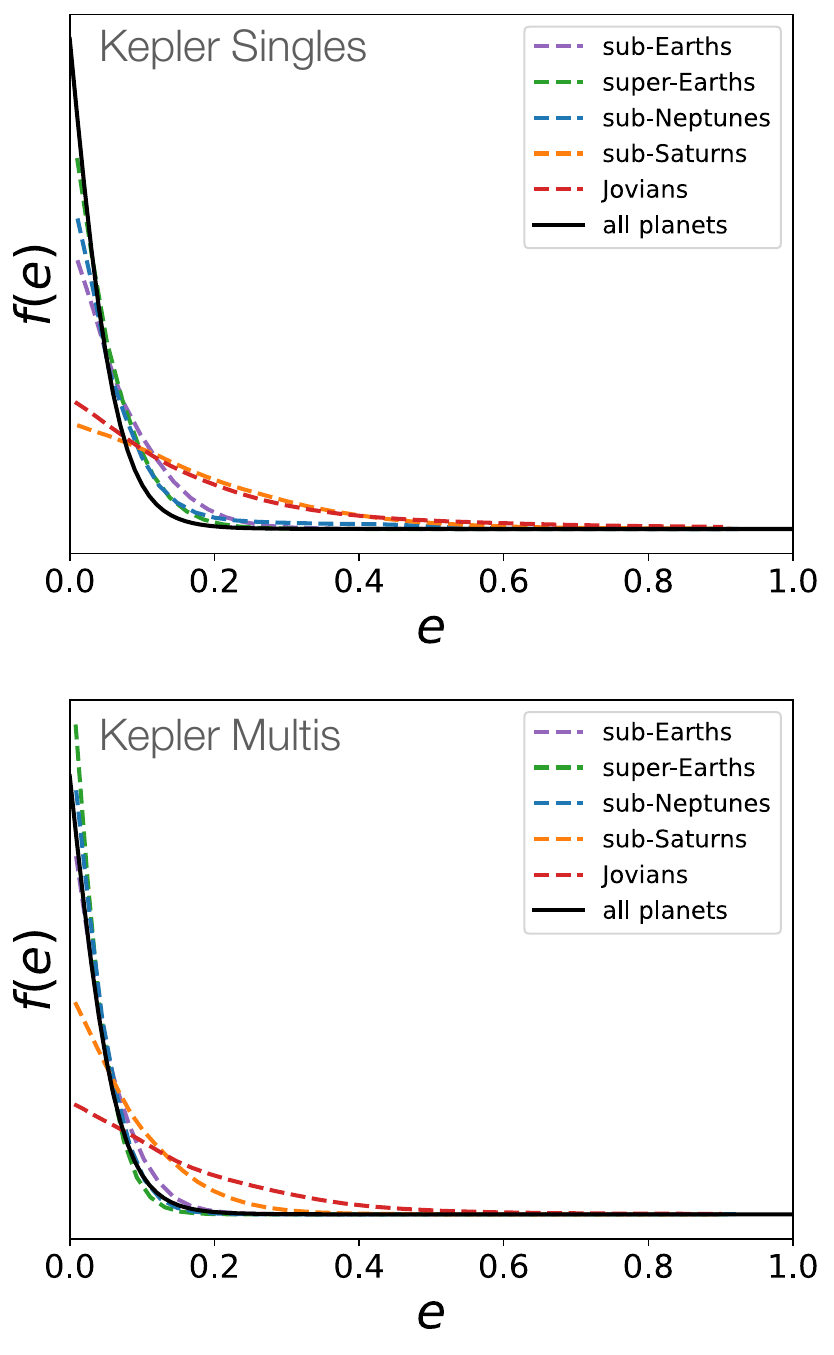}
    \caption{Population-level distribution of eccentricity $f(e)$ inferred using a flexible histogram for Kepler single-transiting systems (top) and multi-transiting systems (bottom). Colorful dashed lines display results for each of five planet size classes, and the solid black line shows the fit for planets of all sizes. All planet size and multiplicity groups exhibit the same basic functional form for $f(e)$, with a mode at $e=0$, falling quasi-exponentially toward zero at $e=1$. The strongest similarity in the shape of the distribution exists between the two large planet classes (sub-Saturns and Jovians), and likewise for the three small planet classes (sub-Earths, super-Earths, and sub-Neptunes).}
    \label{fig:template-dist}
\end{figure}

We found a remarkable similarity in the morphology of $f(e)$ across planet sizes and system multiplicities (Figure \ref{fig:template-dist}). For each of the planet classes, $f(e)$ had a mode at $e=0$ and declined quasi-exponentially and nearly monotonically toward zero at $e=1$. All non-monotonic behavior was at or below the $10^{-2}$ level in the tails of the distribution. We attributed these fluctuations to the highly flexible model over-fitting noise in the low-probability regions of parameter space. We therefore imposed monotonicity on the PDF by extending a shallowly sloped tangent line past the first local minimum in each curve; the slope of the tangent was determined as $\Delta \alpha/ \Delta e$ between the two histogram bins immediately preceding the local minimum. So, the shape of the empirical $f(e)$ for $e$ values lower than the first local minimum are derived directly from the regularized histogram model, while $f(e)$ for $e$ values higher than the first local minimum are assumed to follow a log-linear dependence. These modifications greatly simplified the form of the empirical PDF with only a slight impact on the integrated probabilities.

Having confirmed that the distribution shape is consistent across sub-populations, we constructed an empirical template PDF $f_0$ by fitting our non-parametric model to the full \numplanets planet sample. We then interpolated $f_0$ to a uniform grid of 100 evenly spaced bins in eccentricity ($\Delta e = 0.01$). Finally, we enforced monotonicity as above, applied a small smoothing correction using a univariate spline, and re-normalized the resultant PDF.

\begin{table}
\centering
\caption{Empirical eccentricity distribution}
\begin{tabular}{ll}
$e$ & $\ln f(e)$ \\
\midrule
$0.00$ & $2.7688$ \\
$0.01$ & $2.6319$ \\
$0.02$ & $2.4794$ \\
\vdots & \vdots \\
$0.99$ & $-3.2198$ \\
$1.00$ & $-3.2205$ \\
\bottomrule
\end{tabular}

\addtabletext{Empirical eccentricity distribution $f(e)$ derived from our flexible histogram model. Values of $e$ and $f(e)$ have been interpolated to an even grid with spacing $\Delta e = 0.01$. Only part of the table is shown here for illustrative purposes; the full machine-readable table is available online.}
\label{tab:empirical-template}
\end{table}

When fitting the eccentricity distribution of various sub-populations using the empirical template, we transformed the template PDF $f_0$ as
\begin{equation}
    \ln f_E(e;\nu,h) = \nu \left(\ln f_0(he)\right) + (1-\nu) \left(\ln f_0(0)\right)
\end{equation}
where $\nu$ and $h$ are positive coefficients controlling the scaling in the vertical and horizontal directions, respectively. The resultant distribution was then numerically re-normalized. When the template distribution was compressed in the horizontal direction (i.e. $h > 1$), the tail values were extrapolated using the constant value at $f_0(1)$. To first order, $\nu$ sets the heaviness of the distribution tail (by scaling logarithmically in the vertical direction) and $h$ sets the width of the distribution core (by scaling linearly in the horizontal direction). However, the correspondence is not one-to-one due to the interplay of normalization and scaling. The coefficients $\nu$ and $h$ are strongly anti-correlated. So, when fitting for these quantities we add a regularization term to the log-likelihood $\ln \mathcal{L} = -(\nu-h)^2$ which penalizes large reciprocal values of $\nu$ and $h$ which would cancel each other out.

\begin{SCfigure*}
\centering
\includegraphics[width=0.65\textwidth]{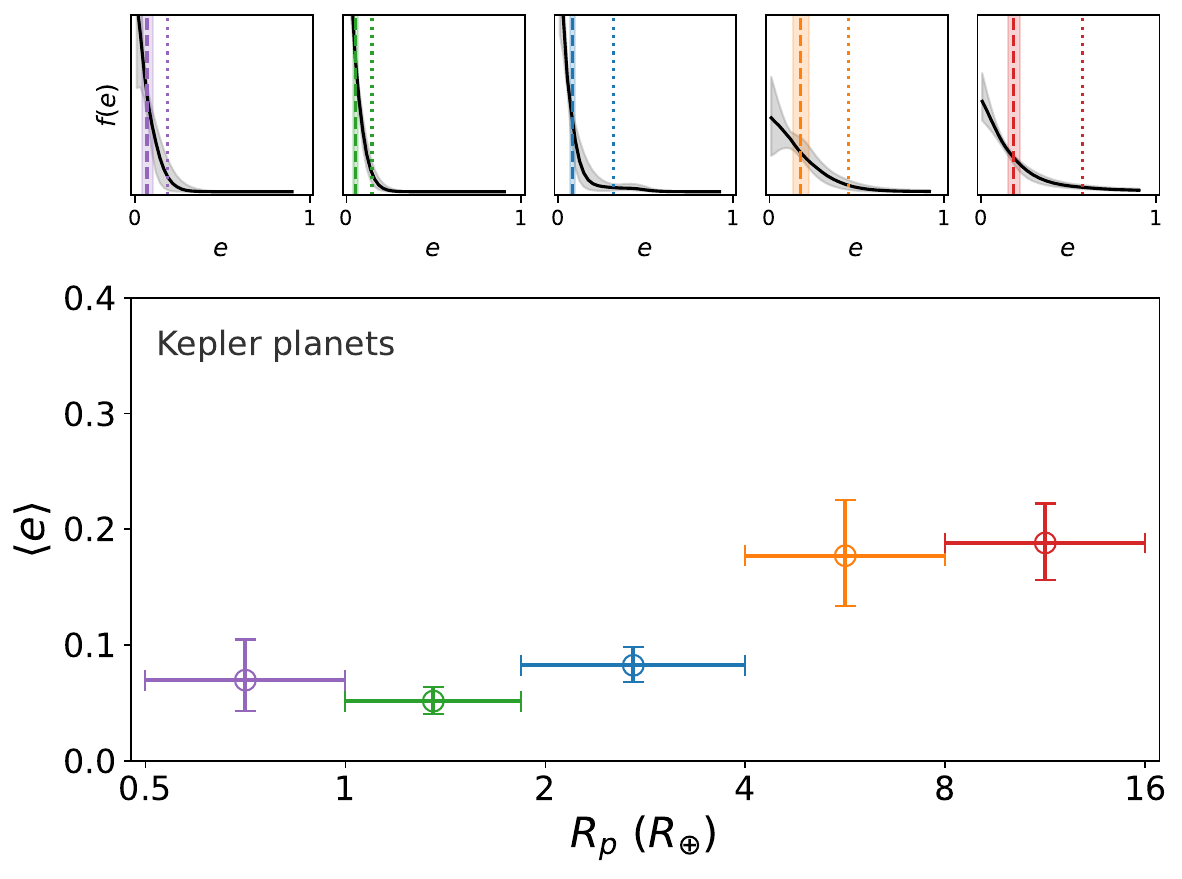}
\caption{Demonstration of our data compression procedure for summarizing population distributions $f(e)$ as $\e$. \textit{Top}: population distributions $f(e)$ for each of five planet size classes. The black line indicates the median inferred $f(e)$ and the grey shaded region indicates the $16^{\rm th}$ to $84^{\rm th}$ percentile confidence interval. Colored dashed (dotted) lines indicate the mean ($95^{\rm th}$ percentile) of $f(e)$. The colored shaded region indicates the 68\% confidence interval on the mean. \textit{Bottom}: Mean eccentricity $\e$ for each planet size class derived by computing the mean and 68\% confidence interval for each sub-population f(e). There is a clear relationship between increased planet size and elevated mean eccentricity.}
\label{fig:marginalization-demo}
\end{SCfigure*}

\begin{figure*}
\centering
\includegraphics[width=0.8\textwidth]{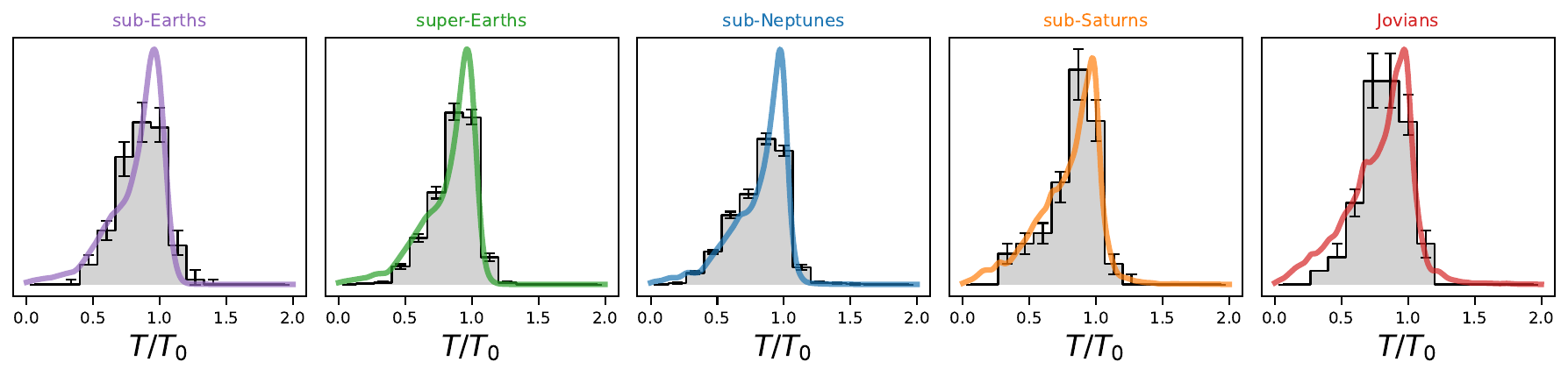}
\caption{Distribution of observed transit duration ratios $\{T/T_0\}_{\rm obs}$ (grey histograms) compared to duration ratios predicted by our best-fit empirical model $\{T/T_0\}_{\rm mod}$ (colored lines) for each of five planet size classes. Error bars on each histogram bin are based on the $16^{\rm, th}$ and $84^{\rm th}$ percentiles of 1000 bootstrap trials. Model predictions have been filtered to account for detection biases in the Kepler pipeline. $\{T/T_0\}_{\rm obs}$ and $\{T/T_0\}_{\rm mod}$ are in good agreement for all five size classes.}
\label{fig:duration-ratio-check}
\end{figure*}

\subsection*{Comparison to parametric distributions}

Previous investigations have modeled the exoplanet eccentricity distribution as a Rayleigh $f_{\rm R}$, half-Gaussian $f_{\rm G}$, and Beta distributions $f_{\rm B}$. These are:

\begin{eqnarray}
   f_{\rm R}(e;\sigma) & = &\frac{e}{\sigma^2}\exp\left(-e^2/2\sigma^2\right), \\
   f_{\rm G}(e;\sigma) &= & \frac{\sqrt{2}}{\sigma\sqrt{\pi}}\exp\left(-e^2/2\sigma^2\right), \\
   f_{\rm B}(e;a,b) &=& \frac{\Gamma(a+b)e^{a-1}(1-e)^{b-1}}{\Gamma(a)\Gamma(b)}.
\end{eqnarray}

In our preliminary analysis, we found that inference using these three distributions was sensitive to the presence of outliers. In general, it is dangerous to fit one- or two- parameter models in the presence of unknown outliers, as a few anomalous objects (or samples) can bias inference for the bulk of the population. We therefore modified each of these distributions by adding a small constant term $\epsilon$ such that $f(e) \rightarrow f(e) + \epsilon$, after which the resultant distribution was re-normalized. We experimented with different values of $\epsilon$ and found that between $\epsilon\approx10^{-9}-10^{-3}$ the results were not sensitive to exact choice of $\epsilon$. We therefore set $\epsilon = 10^{-6}$ as our nominal value. We found that models which include $\epsilon$ are more robust against high-$e$ outliers and produce somewhat lower inferences of mean eccentricity $\e$ overall. We interpret this result as $\epsilon$ providing an ad-hoc correction that accounts for both noise and the heavy tails of the true underlying distribution. We note that results were qualitatively consistent -- albeit noisier -- when the models were fit without the $\epsilon$ correction.

\begin{figure*}
    \centering
    \includegraphics[width=0.95\textwidth]{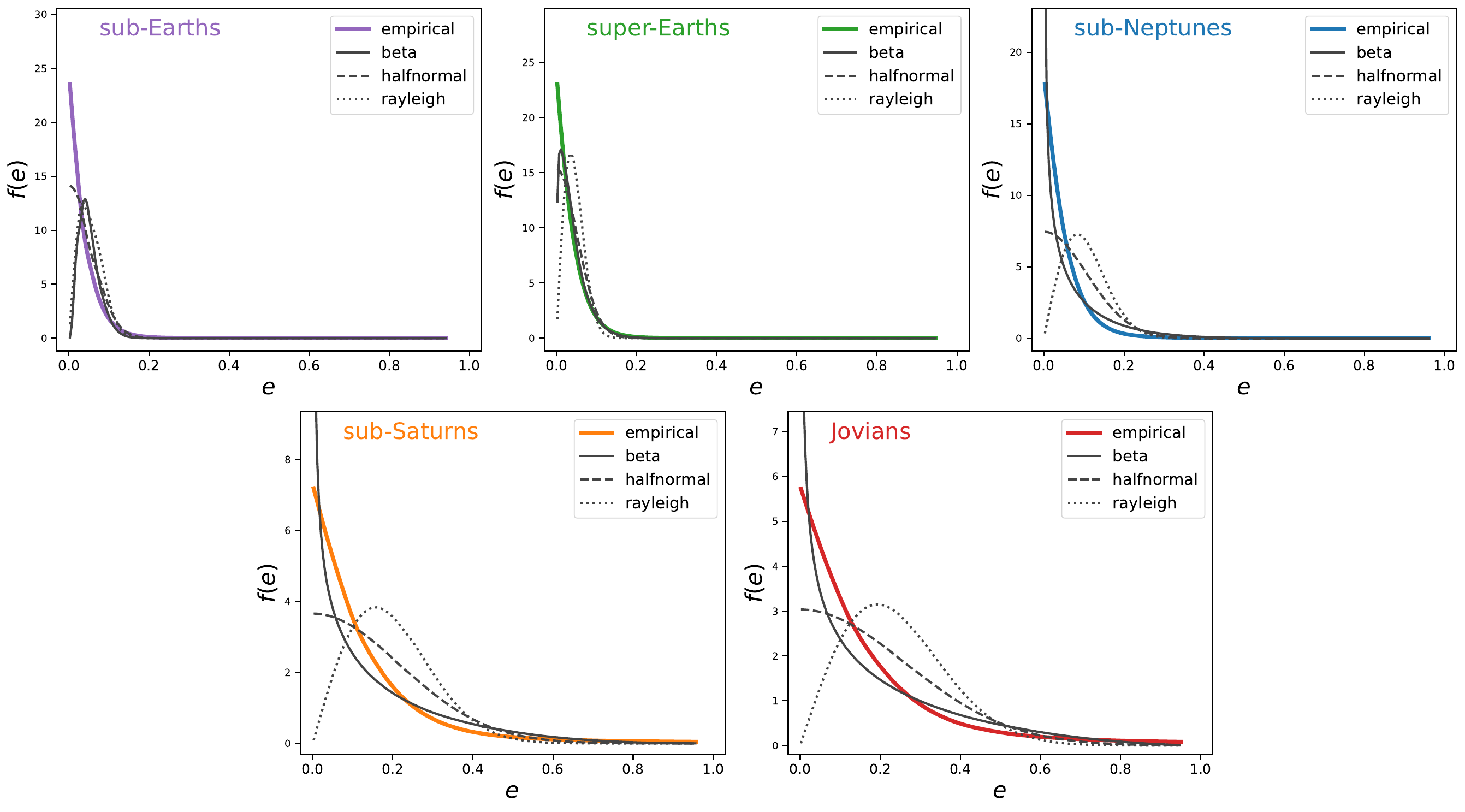}
    \caption{Maximum likelihood posterior inferences of the eccentricity distribution $f(e)$ assuming different parametric models for each of our five plannet size classes. The thick colored line shows our preferred empirical model derived from a flexible histogram. Beta (solid line), half-Gaussian (dashed line), and Rayleigh (dotted line) models are shown in grey. The vertical axis for each panel has been scaled to the maximum value of the empirical distribution. For all five planet classes, the Rayleigh distribution is the clear outlier among the four models, with a forced endpoint value $f(0)=0$; the three other models all have their mode at $f(0)$.}
    \label{fig:parametric-model-compare-pdfs}
\end{figure*}

Figure \ref{fig:parametric-model-compare-pdfs} shows the maximum likelihood inference of $f(e)$ obtained using the beta, half-Gaussian, and Rayleigh distributions, as well as our scaled empirical model for each of the five planet size classes. Median parameter values and uncertainties for the empirical, beta, and half-Gaussian distributions are presented in Table \ref{tab:parameter-inferences}. The Rayleigh distribution provides a poor description of the data (see below for discussion) and so is not included in the table.

\begin{table*}
\centering
\caption{Inferred Eccentricity Distribution Model Parameters}
\begin{tabular}{l|l|l|l|l}
Population & Size Range ($R_{\oplus}$) & Empirical & Beta & Half-Gaussian\\
\midrule

Jovians & $8-16$ & $\nu=0.68^{+0.31}_{-0.14}$, $h=0.55^{+0.27}_{-0.21}$ & $a=0.58^{+0.19}_{-0.14}$, $b=2.62^{+0.89}_{-0.66}$ & $\sigma=0.26^{+0.03}_{-0.03}$\\

sub-Saturns & $4-8$ & $\nu=0.80^{+0.25}_{-0.16}$, $h=0.60^{+0.29}_{-0.20}$ & $a=0.56^{+0.26}_{-0.15}$, $b=3.33^{+1.45}_{-1.01}$ & $\sigma=0.16^{+0.02}_{-0.02}$\\

sub-Neptunes & $R_{\rm gap}-4$ & $\nu=1.01^{+0.09}_{-0.06}$, $h=1.21^{+0.29}_{-0.23}$ & $a=0.54^{+0.07}_{-0.05}$, $b=7.35^{+1.15}_{-0.96}$ & $\sigma=0.11^{+0.01}_{-0.01}$\\

super-Earths & $1.0 - R_{\rm gap}$ & $\nu=1.60^{+0.34}_{-0.25}$, $h=1.02^{+0.28}_{-0.21}$ & $a=1.40^{+1.17}_{-0.50}$, $b=33.36^{+29.40}_{-11.15}$ & $\sigma=0.05^{+0.01}_{-0.01}$\\

sub-Earths & $0.5 - 1.0$ & $\nu=1.48^{+0.46}_{-0.31}$, $h=1.16^{+0.56}_{-0.39}$ & $a=3.94^{+20.14}_{-2.48}$, $b=75.08^{+393.02}_{-50.16}$ & $\sigma=0.06^{+0.03}_{-0.02}$\\

\bottomrule
\end{tabular}

\addtabletext{Posterior inferences of model parameters for the empirical (Equation \ref{eq:empirical-pdf}), beta, and half-Gaussian distributions. Results are presented for each of our five planet size classes. Quoted values on parameters give the $16^{\rm th}$, $50^{\rm th}$, and $84^{\rm th}$ percentiles. The Rayleigh distribution provides a poor description of the data and so is not included in the table.}
\label{tab:parameter-inferences}
\end{table*}

To quantify the quality of the various model fits, we performed an analysis of the Bayesian Information Criterion \citep[BIC,][]{Schwarz1978-BIC} and the Akaike Information Criterion \citep[AIC][]{Akaike1974-AIC}. Based on these quantities, we find that our empirical distribution provides as good as or a better match to the data for planets of all sizes. On average, the beta distribution generally performs comparably to our empirical distribution ($\Delta\rm BIC$ and $\Delta\rm AIC \leq 6$), while the half-normal distribution is modestly disfavored ($\Delta\rm BIC$ and $\Delta\rm AIC \approx 10$) and the Rayleigh distribution is strongly disfavored ($\Delta\rm BIC$ and $\Delta\rm AIC \approx 25$). The greatest differences are observed for super-Earths and sub-Neptunes, the two most populous size classes, which favor either the empirical or beta distributions. For the sub-Saturns and Jovians, the Rayleigh distribution can be confidently ruled out, while the other three distributions are statistically tied. No distribution outperforms all of the others under all circumstances. We interpret the lack of a clear distinction between the empirical, half-normal, and beta distributions as due to either the small number of large planets ($R_p > 4 \Re$) or the near-zero eccentricities for small planets ($R_p < 4 \Re$), which make subtle differences between the distributions difficult to detect. Only the Rayleigh distribution, which does not permit a mode at $e=0$, can be confidently ruled out by the data.

To further quantify the quality of the model fits, we compared observed values of transit duration ratio $T/T_0$ obtained from Kepler data with forward modeled values of $T/T_0$ derived from each of the four eccentricity distributions. Our forward modeling procedure was as follows. For each distribution and planet size class, we drew parameter values within the uncertainties reported in Table \ref{tab:parameter-inferences}. We then calculated the implied eccentricity distribution, from which we drew $N$ samples of $e$, where $N$ is the number of planets. We combined these samples of $e$ with samples of $\omega$ drawn from a uniform distribution $\omega\sim\mathcal{U}(0,2\pi)$ and samples of $b$ drawn from $b \sim (1-b^2)^{1/4}$, as derived in \citealp{KippingSandford2016}. From these $\{e,\omega,b\}$ we calculated the duration ratio following Equation \ref{eq:photo-ecc-SI}.

We added a Gaussian measurement uncertainty $\sigma(T/T_0)$ equal to the median uncertainty on $T$ for all observed Kepler planets in a given planet size bin. To account for sampling noise in Gaia-Kepler data and in our inferred eccentricity distribution, we bootstrapped the above procedure for 1000 trials. These bootstrap tests generated a set of model duration ratios $\{T/T_0\}_{\rm mod}$ which we then compared to the observed duration ratios $\{T/T_0\}_{\rm obs}$ derived from our transit modeling of Kepler photometry. This comparison was quantified by computing the Kolmogorov-Smirnov \citep{Kolmogorov1933, Smirnov1939} and Anderson-Darling \citep{AndersonDarling1954} test statistics to determine if $\{T/T_0\}_{\rm mod}$ and $\{T/T_0\}_{\rm obs}$ were drawn from the same underlying distribution. The advantage of working with duration ratios rather than with eccentricity is that transit duration is a directly observable quantity with small fractional uncertainties $(\sigma_T \sim 5\%)$. 

\begin{figure*}
    \centering \includegraphics[width=0.95\textwidth]{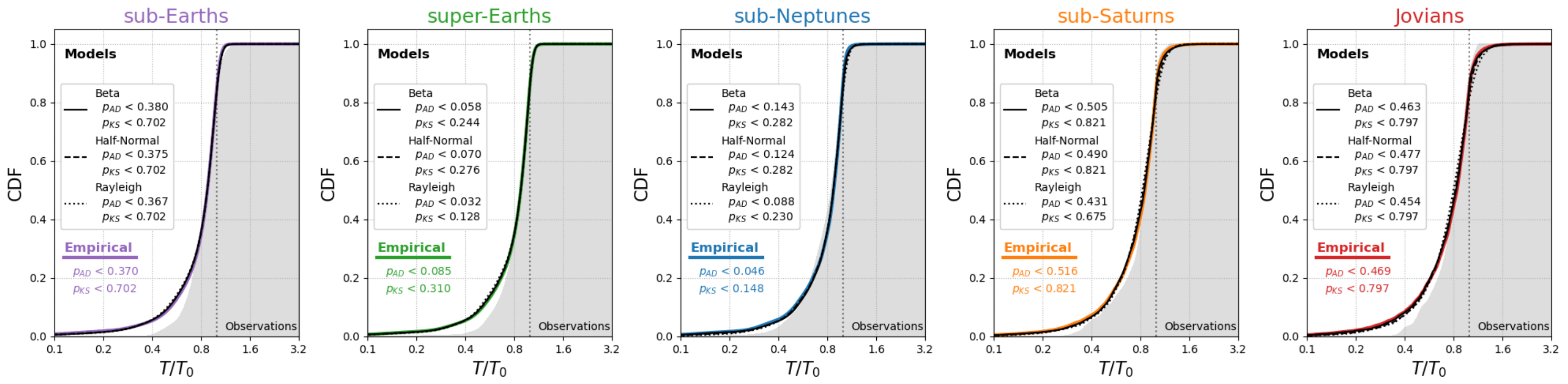}
    \caption{Cumulative distribution functions (CDF) for transit duration ratio $T/T_0$ for each of five planet size classes. Observations from our transit modeling of Kepler photometry are shown as a grey shaded region. Forward-modeled values of $T/T_0$ based on our empirical model are shown as solid colored lines. Forward modeled values based on the beta, half-Gaussian, and Rayleigh models are shown as solid, dashed, and dotted black lines, respectively. For each model, the $90^{\rm th}$ percentile upper limit on the Anderson-Darling (AD) and Kolmogorov Smirnov (KS) test statistic p-value are displayed on the chart. For the three sub-Neptunes, the empirical and beta models are preferred, while the half-Normal distribution is modestly disfavored and the Rayleigh distribution is strongly ruled out. For all other planet classes, there is no apparent preference between models based on the AD and KS tests.}
    \label{fig:cdf_comparison}
\end{figure*}

Figure \ref{fig:cdf_comparison} shows the median cumulative distribution function for the various $\{T/T_0\}_{\rm mod}$ as compared to $\{T/T_0\}_{\rm obs}$ for each of the five planet size classes. The figure also displays $90^{\rm th}$ percentile upper limits of the KS and AD test statistic p-values derived from our bootstrap trials. We find that based on these quantities, all four distributions (empirical, beta, half-Gaussian, Rayleigh) are formally acceptable in the majority of cases. This agreement in terms of transit duration ratios underscores the necessity of applying a hierarchical Bayesian analysis. Whereas a duration ratio approach obscures covariances between $b$, $e$, and $\omega$, our full hierarchical approach accounts for these covariances, allowing us to confidently rule out the Rayleigh distribution and detect the distribution mode at $e=0$.

As an additional check, we computed the Kullback-Liebler divergence \citep{KullbackLeibler1951}, which quantifies the information loss when approximating the $\{T/T_0\}_{\rm obs}$ by each of the four parametric models. We find the the KL divergence is statistically indistinguishable between the four parametric models, although the Rayleigh distribution tends to perform worst at a sub-significant level. We interpret the lack of a distinction between models as a byproduct of the generally low eccentricities of real planets. For planets of all sizes, $\e \lesssim 0.3$, which translates into a relative change on $T/T_0 \lesssim 5\%$; indeed, 99\% of samples of $\{T/T_0\}_{\rm obs}$ fall between 0.4--1.3. Consequently, subtle differences in the distribution of eccentricity will be difficult to detect in the observable space $T/T_0$. Nevertheless, it is reassuring to see that the models which are preferred by the AIC and BIC (evaluated directly on eccentricity) are also permitted by the KL divergence, AD test, and KS test evaluated on duration ratios.

Based on the analysis above, we conclude that both the beta distribution and our empirical distribution provide good descriptions of exoplanet eccentricities. The beta distribution has the advantage of being simpler to implement, while our empirical distribution has the advantage of avoiding a singularity at $e=0$. The half-Gaussian distribution is modestly disfavored, although it provides a passable single-parameter description of the data after being modified to account for potential outliers in the tail (i.e. $f(e) \rightarrow f(e) + \epsilon$). The Rayleigh distribution provides a poor description of exoplanet eccentricities and should not be used.

\section*{Estimating the significance of the eccentricity peak in the radius valley}

\begin{figure*}
    \centering \includegraphics[width=0.95\textwidth]{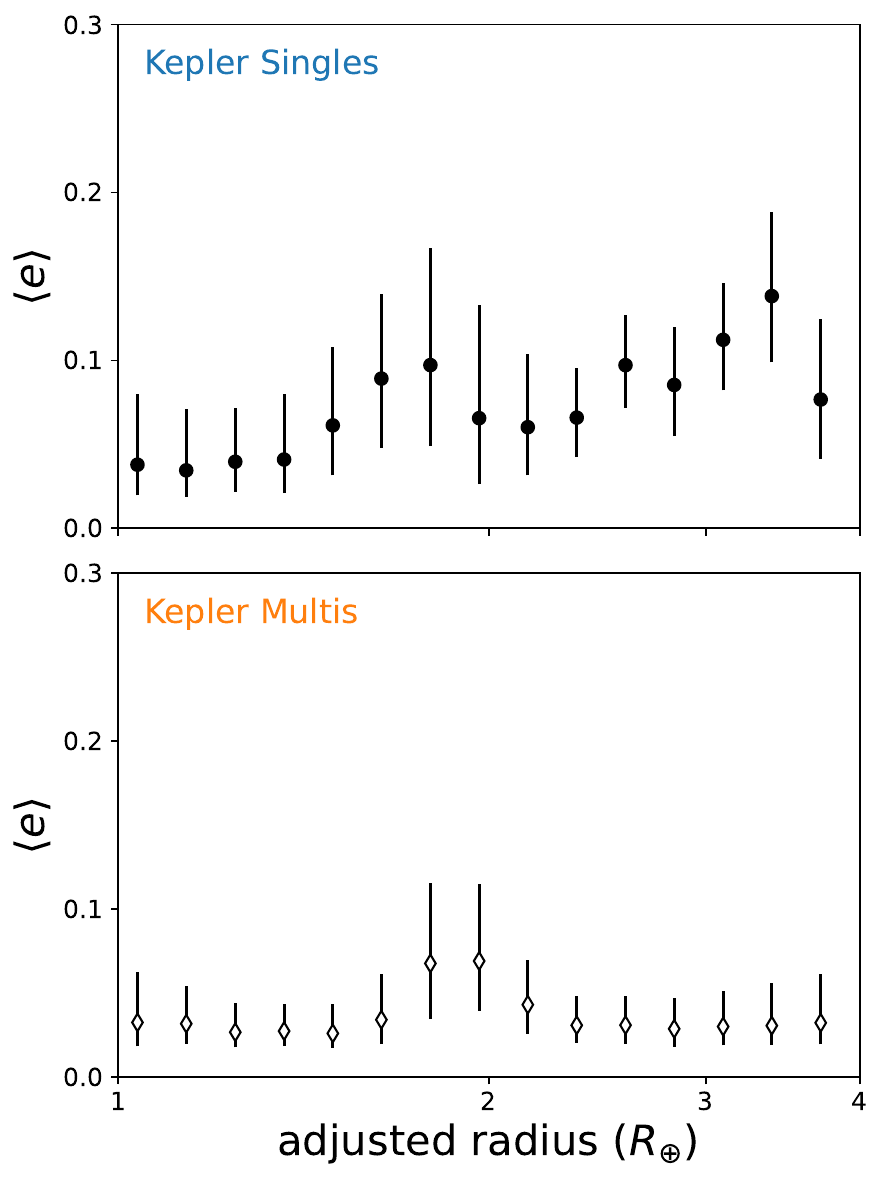}
    \caption{Mean eccentricity as a function of adjusted planet radius for planets in single-transiting (top) and multi-transiting (bottom) systems. Smooth blue/orange curves denote the posterior fit and 68\% confidence interval for a logistic sigmoid + Gaussian model, which we use to estimate the significance of the peak at $2.1\sigma$.}
    \label{fig:ecc-rp10-model}
\end{figure*}

We calculate the significance of the eccentricity peak in the radius valley as follows. First, we determine $\e$ as a function of adjusted planet radius $R_p'$ using our hierarchical model. We then fit the $\e$-$R_p'$ curve with a logistic sigmoid + Gaussian
\begin{equation}
    f(x) = B + \frac{L}{1+e^{-k(x-x_t)}} + Ae^{-\frac{1}{2}(x-x_p)^2/s^2}.
\end{equation}
Here, $x\equiv\log R_p'$, $B$ encodes the baseline eccentricity, $L$ encodes the sigmoid normalization, $\{x_t, k\}$ encodes the location and rate of change from low to high $e$, and $\{x_p, s, A\}$ encodes the location, width, and amplitude of the Gaussian peak. In physical terms, $e_{\rm low} = B$ and $e_{\rm peak} = A + B$. See Figure \ref{fig:ecc-rp10-model}.

When fitting the sigmoid + Gaussian model, we allow the Gaussian amplitude parameter to explore both positive and negative values. For singles, $A > 0$ for 73\% of samples (a $1.1\sigma$ measurement); for multis,  $A > 0$ for 89\% of samples (a $1.6\sigma$ measurement). These two measurements are independent. Thus, our confidence that the peak in the valley is positive can be calculated as $1 - P(A_{\rm singles} \leq 0)*(P_{\rm multis} \leq 0) = 0.97$, a $2.1\sigma$ detection.

\section*{Testing for insensitivity to modeling choices}

\subsection*{Do the results depend on population definition?} We experimented by setting narrower cutoffs for planetary orbital period, finding that our results remained consistent when the model was refit with either short period ($P < 10$ day) or long period ($P > 30$ day) planets removed from the sample. We also experimented with setting stricter cutoffs on stellar radius and metallicity, again finding that our results remained consistent. We experimented by setting various stellar magnitude thresholds, finding results qualitatively unchanged whether such a threshold was used or not. 
\subsection*{Do the results depend on planetary detection biases?} We re-ran the analysis both with and without accounting for geometric effects that favor detection of planets on eccentric orbits \citep{Barnes2007, Burke2008} and at low impact parameters \citep{KippingSandford2016}. The inclusion of such effects made virtually no change to our inferred population eccentricity distributions. 
\subsection*{Do the results depend on the parametric form of the eccentricity distribution?} 
\begin{figure*}
\centering
\includegraphics[width=0.95\textwidth]{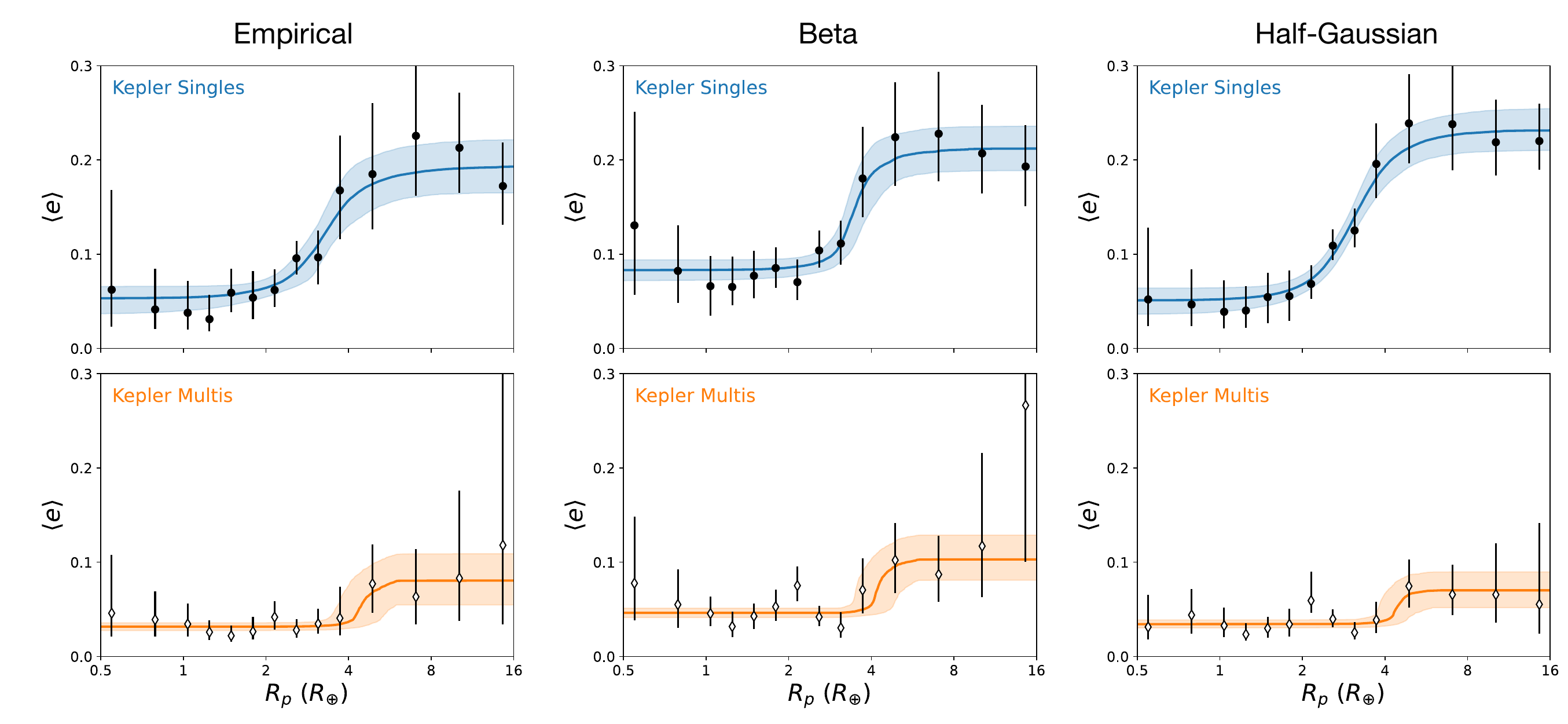}
\caption{Relationship between mean eccentricity $\e$ and planet size $R_p$ derived using our empirical distribution (left) vs a beta distribution (middle) or half-Gaussian distribution (right). Results are qualitatively consistent regardless of which model is adopted. While there is some variation in the amplitude and significance of features in the $\e-R_p$ curve, our conclusions are not strongly impacted by the choice of model.}
\label{fig:compare-ecc-rp}
\end{figure*}

We ran the analysis using the half-normal, beta, and empirical distributions, finding consistent results across all iterations (Figure \ref{fig:compare-ecc-rp}).
\subsection*{Do the results depend on the method of radius binning?} We ran the analysis using (1) a standard likelihood and top-hat binning, and (2) a weighted likelihood and weighted binning; we found found qualitatively identical results when using both the simple and the more complicated method. For the top-hat binning, we performed a standard boostrap test \citep{Effron1979}, and for the weighted likelihood binning we performed a weighted likelihood bootstrap \cite{NewtonRafferty1994} to rule out sampling noise as the source of the trends. Uncertainties scaled as expected during these tests and posterior inferences remained consistent across iterations.

\section*{Injection-and-recovery tests}

We performed a suite of injection-and-recovery experiments to validate our transit fitting pipeline and hierarchical eccentricity analysis. Our goal was to investigate the influence of (1) transit shape and signal-to-noise ratio, (2) correlated noise, and (3) TTVs on posterior inferences of population eccentricity. Our procedure is as follows.

First, we simulated a population of 1302 planets with properties designed to provide a rough facsimile of the real Kepler sample. For the purpose of this experiment, it was not essential to exactly match the real distribution, but rather to provide support over parameters spanned by the Kepler sample. To construct our simulated catalog of planets, we selected all single-transiting systems in Kepler DR25 \citep{Thompson2018} with reported periods $P = 1-100$ days and size ratios $R_p/R_\star < 0.2$. We then eliminated from our sample all planets with unreliable impact parameters following the criteria of \citealp{Petigura2020}, after which each planet was assigned a synthetic impact parameter $b$ drawn from $b \sim (1-b^2)^{1/4}$, as suggested by \citealp{KippingSandford2016}. Stellar parameters were drawn from physically motivated Gaussian distributions based on Solar values. For synthetic planets with $R_p < 4 \Re$, we drew eccentricity $e$ from a half-normal distribution with $\sigma(e) = 0.03$, and for synthetic planets with $R_p \geq 4 \Re$ we drew $e$ from the Doppler giant planet beta distribution \citep{Kipping2013-beta}, $e\sim\mathcal{B}(a=0.897,b=3.03)$. In this way, our small planet population had $\e \approx 0.023$ and our large planet population had $\e \approx 0.23$. In all cases arguments of pericenter were drawn uniformly. For planets with $a/R_\star < 5$, we enforced $e=0$, as expected from tidal circularization \citep{RasioFord1996, Jackson2008, DawsonJohnson2018}. For each set of simulated parameters, we checked to ensure that $T_{14} > 1$ hr.

\begin{table}
\centering
\caption{Simulated Parameter Distributions for Injection-and-Recovery Tests}
\begin{tabular}{lrc}
Quantity & Distribution & Units \\
\midrule
$P$ & DR25 & days \\
$R_p/R_\star$ & DR25 & - \\
$b$ & $(1-b^2)^{1/4}$ & - \\
$e_{\rm small}$ & $\mathcal{N}(0,0.03)$ & - \\
$e_{\rm large}$ & $\mathcal{B}(0.897,3.03)$ & - \\
$\omega$ & $\mathcal{U}(0, 2\pi)$ & rad \\
\midrule
$u_1$ & $\mathcal{N}(0.4,0.1)$ & - \\
$u_2$ & $\mathcal{N}(0.25,0.1)$ & - \\
$R_{\star}$ & $\mathcal{N}(1.0,0.05)$ & $R_{\odot}$ \\
$\rho_{\star}$ & $\mathcal{N}(1.41,0.07)$ & $\rm g/cm^3$ \\

\bottomrule
\end{tabular}

\addtabletext{Simulated ``ground truth'' distributions for transit and host star parameters. All quantities take their usual values as defined in the main text. In the table above, $\mathcal{U}(\text{low,high})$ indicates a uniform distribution, $\mathcal{N}(\mu,\sigma)$ indicates a normal distribution, and $\mathcal{B}(a,b)$ indicates a beta distribution. All distributions were chosen to produce a rough fascimile of the Kepler population while providing support over the relevant regions of parameter space.}
\label{tab:injected-distributions}
\end{table}

To accurately capture real photometric noise properties, we injected the simulated transits back into the set of single-planet Kepler light curves. To prepare the light curves, we removed the real transit signal using transit parameters drawn from DR25 and TTVs drawn from \citealp{Holczer2016}. While this method may leave minor residual structure at the original transit epochs, we avoided them by injecting the simulated transit with a phase shift of $61.8\%$. For each planet, we simulated TTVs by randomly assigning one of four TTV morphologies: linear (i.e. no TTVs), cubic polynomial, single-component sinusoidal, or boxcar-smoothed white noise; this last case produced smoothly varying TTVs with complex structure. In all cases, we drew TTV amplitudes from a normal distribution $\mathcal{N}(0,T_{14}/3)$ and timescales between $2P$ and $t_{\rm max}-t_{\rm min}$. The incidence of detected TTVs in the real Kepler sample is $\sim10\%$ \citep{Holczer2016}, meaning our method modestly overstates the influence of TTVs, even after accounting for the fraction which possess randomly drawn low amplitudes.

After constructing these synthetic transits lightcurves, we fit the transits using the \alderaan pipeline and a method identical to that which we applied to real data. The only modification to our procedure was to initialize our model close to the known injection values rather than using values in \citealp{Thompson2018} and \citealp{Holczer2016}.

\begin{figure*}
    \centering
    \includegraphics[width=0.95\textwidth]{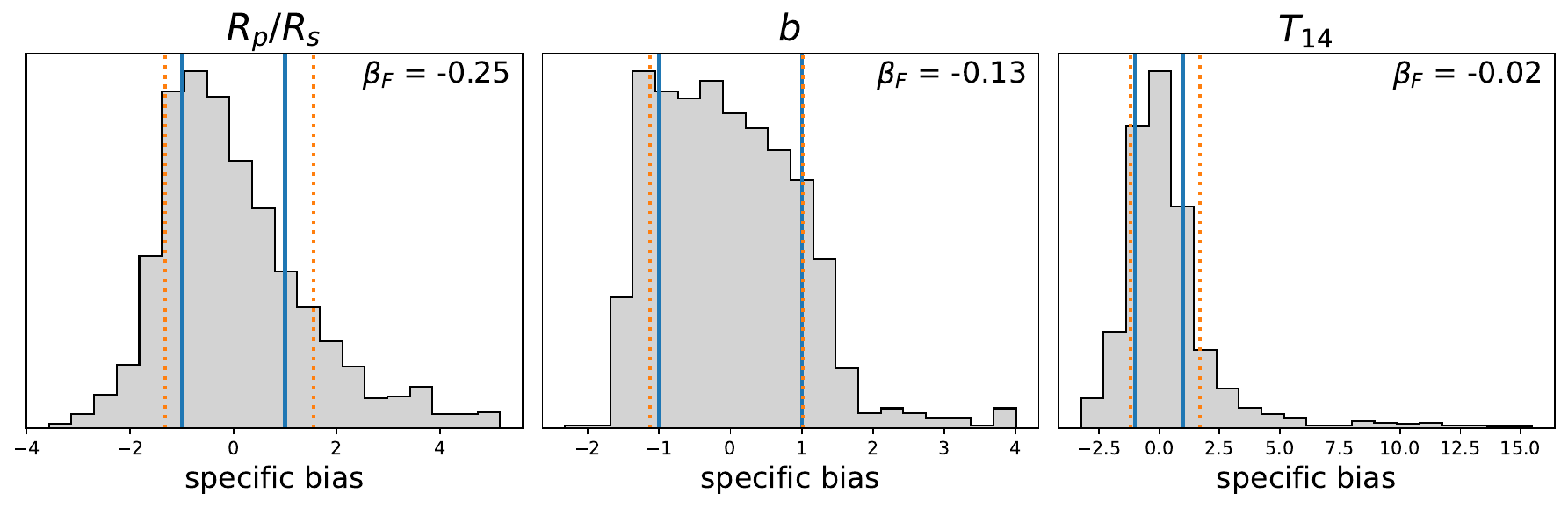}
    \caption{Results of specific bias tests used to validate our injection-and-recovery experiment. Specific bias is calculated from Equation \ref{eq:specific-bias} following \cite{ONeil2019}. Grey histograms show the distribution of specific biases for $R_p/R_\star$ (left), $b$ (middle), and $T_{14}$ (right) for the 1302 synthetic planets. Our recovered 68\% confidence intervals (orange dotted lines) closely match the expected 68\% confidence intervals (blue solid lines), and overall bias is low ($\beta_F \approx 0$), indicating good agreement between injected and recovered results.}
    \label{fig:injection-recovery-coverage}
\end{figure*}

Each injection-and-recovery test produced a set of posterior samples of transit parameters. We assessed the accuracy of these posterior distributions by comparing them to the injected values. We wish to test (1) whether the set of samples are biased, i.e. are their medians systemically higher or lower than the true value? and (2) whether their spread is correct, i.e. do the right number of samples fall above/below the true value? To do so, we followed \cite{ONeil2019} and computed the specific bias 
\begin{equation}\label{eq:specific-bias}
\beta_i = \frac{\hat{\theta}_i - \theta_{i,\rm true}}{\sigma_i}
\end{equation}
where $i$ is an index labeling each of the 1302 injections, $\hat{\theta_i}$ is the median of the fit, $\theta_{i,\rm true}$ is the injected value, and $\sigma_i$ is the 68\% credible interval. Figure~\ref{fig:injection-recovery-coverage} shows that, for the key transit parameters, $\beta$ is centered around zero (no bias) and the spread is correct $\sim$68\% of $\beta_i$ lie between -1 and 1.

\begin{SCfigure*}
    \centering
    \includegraphics[width=0.65\textwidth]{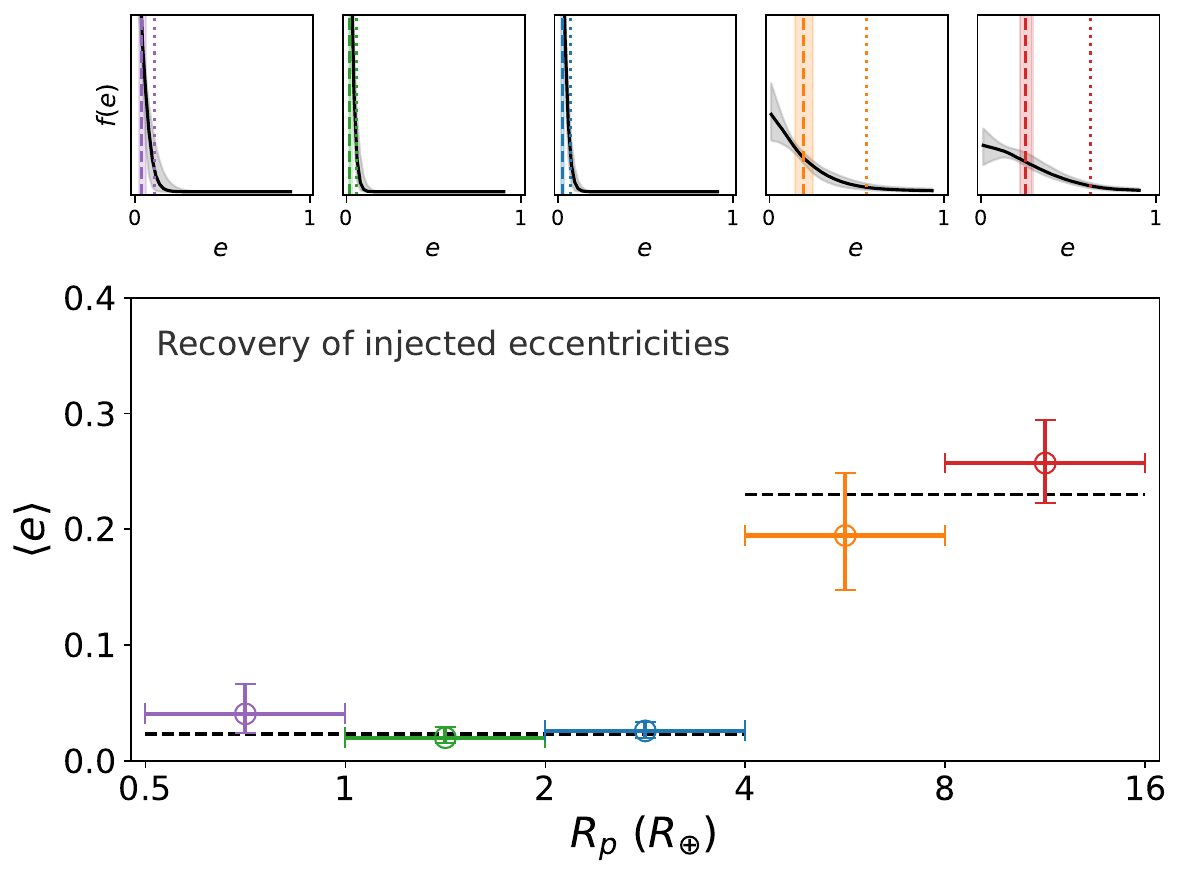}
    \caption{Identical to Figure~\ref{fig:marginalization-demo} except it shows the successful recovery of injected eccentricities. The dashed black line indicates the ground-truth injected mean eccentricity: $\e = 0.023$ for small planets and $\e = 0.23$ for large planets. The recovered $\e$ values are consistent with the injected values within to their reported uncertainties.}
    \label{fig:injection-recovery-eccentricity}
\end{SCfigure*}

\begin{figure*}
    \centering
    \includegraphics[width=0.95\textwidth]{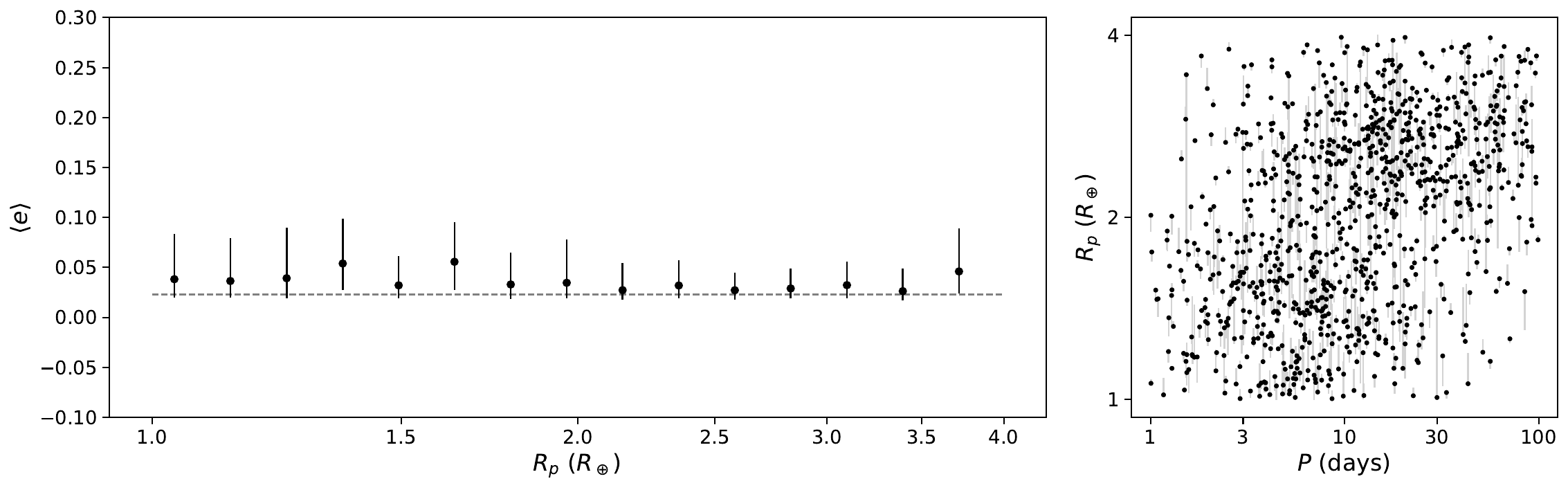}
    \caption{Recovered $\e-R_p$ relationship for ${\sim}~900$ simulated planets between $R_p=1-4\Re$ and $P = 1-100$ days. \textit{Left}: The dashed black line indicates the ground-truth $\e$=0.023. \textit{Right}: Due to measurement uncertainty, some ``observed'' radii (black points) are different than their true radii; grey lines indicate movement from ``ground truth'' to ``observed'' positions.}
    \label{fig:injection-test-gap00}
\end{figure*}

Following the procedure described in the main text, we fed the posteriors into our hierarchical eccentricity analysis and derived the eccentricity distributions of various size classes of planets using our non-parametric method. Figure~\ref{fig:injection-recovery-eccentricity} is analogous to Figure~\ref{fig:marginalization-demo} and shows that our inferences of $\e$ and its uncertainty are consistent with the injected values. In addition, it shows that we correctly derive the shapes of the injected distributions. In addition, we include a higher resolution view of the recovered eccentricities for small planets in Figure~\ref{fig:injection-test-gap00}.

\begin{figure*}
    \centering
    \includegraphics[width=0.95\textwidth]{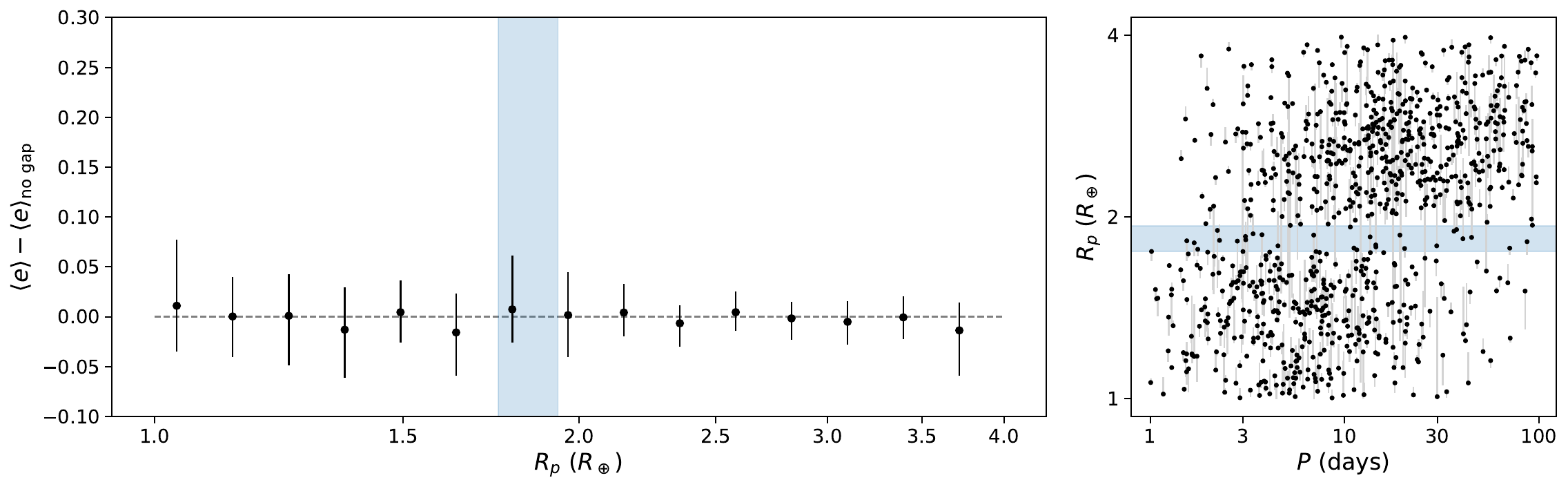}
    \caption{Recovered $\e - R_p$ relationship for injection test with a simulated radius gap between $R_p = 1.75 - 1.93 \Re$. The blue shaded region indicates the region from which all ground truth planets were removed. \textit{Left}: Difference in inferred $\e$ for this simulation vs. the ``no gap'' simulation shown in Figure \ref{fig:injection-test-gap00}. \textit{Right}: Due to measurement uncertainty, some ``observed'' radii (black points) fall within the gap.}
    \label{fig:injection-test-gap10}
\end{figure*}

\begin{figure*}
    \centering
    \includegraphics[width=0.95\textwidth]{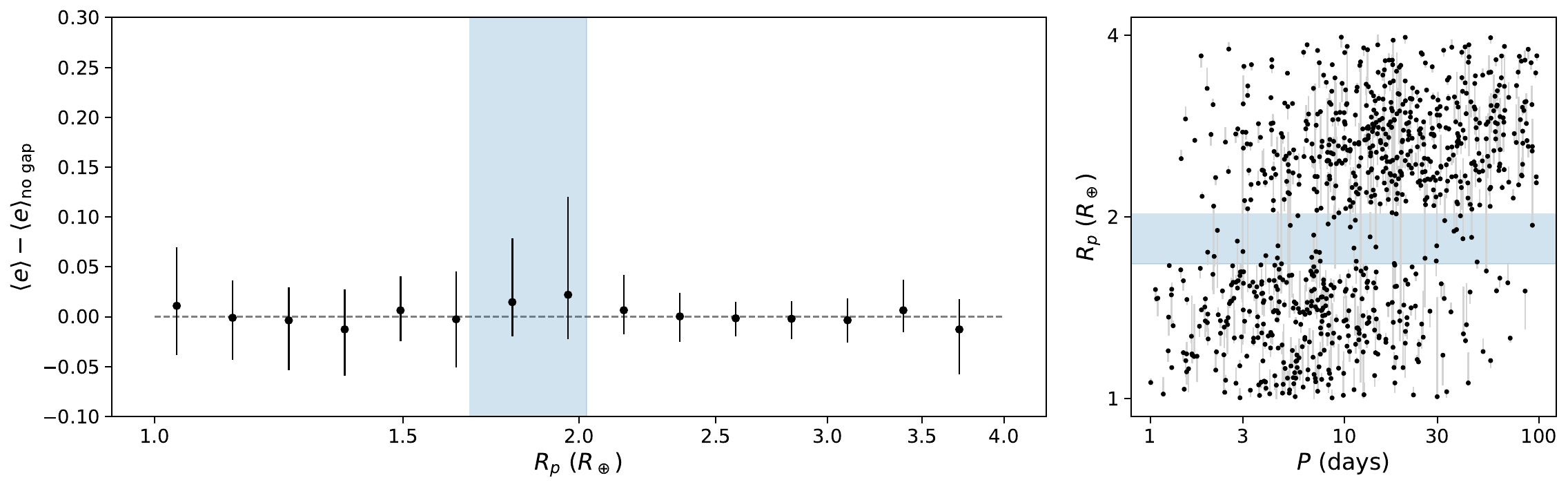}
    \caption{Same as Figure \ref{fig:injection-test-gap10}, but with a simulated radius gap between $R_p = 1.67 - 2.02 \Re$.}
    \label{fig:injection-test-gap20}
\end{figure*}

\begin{figure*}
    \centering
    \includegraphics[width=0.95\textwidth]{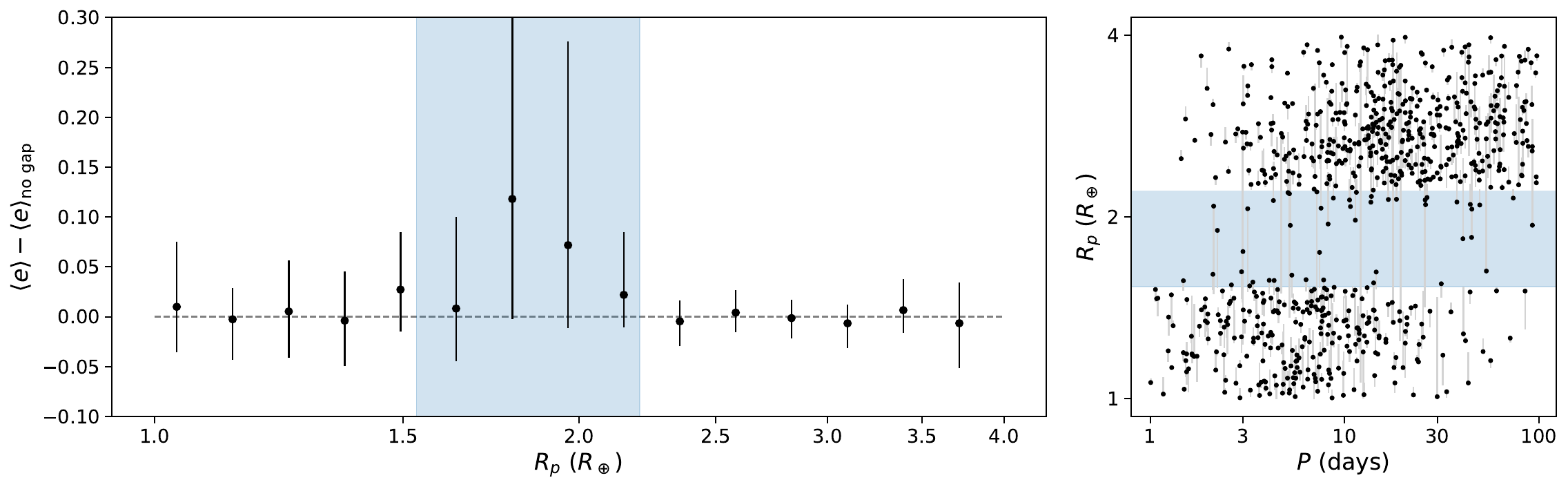}
    \caption{Same as Figure \ref{fig:injection-test-gap10}, but with a simulated radius gap between $R_p = 1.53 - 2.21 \Re$.}
    \label{fig:injection-test-gap40}
\end{figure*}

To more closely investigate the eccentricity peak in the radius valley, we repeated the hierarchical analysis after removing all planets whose injected radii were between 1.75--1.93~\Re. In other words, we simulated a completely depopulated radius gap with a fractional $10\%$ width centered on a nominal $R_p=1.84$~\Re (see \citealp{Petigura2022}). Because photometric noise results in errors on \Rp/\Rstar, our ``observed'' radius valley had a lower occurrence rate density but was not completely depopulated. We inferred $\e$ for each radius bin via our hierarchical analysis and compared this $\e-R_p$ relationship to an $\e-R_p$ relationship inferred without removing any planets. We found no statistically significant change in $\e$ for $R_p$ bins fully or partially within the simulated radius valley (Figure \ref{fig:injection-test-gap10}). Although $\e$ in the radius valley is indeed slightly raised, the commensurate increase in uncertainty is larger both in an absolute sense and in comparison to the uncertainty observed in real data. We repeated this experiment using gaps of width 20\% (Figure \ref{fig:injection-test-gap20}) and 40\% (Figure \ref{fig:injection-test-gap40}). For all three trials, we find that we cannot rule out a physical origin for the eccentricity peak in the radius valley.

\section*{Software}\label{sec:software}
We acknowledge the teams who developed following open-source software packages used in this work: \texttt{astropy} \citep{astropy:2018}, \texttt{batman} \citep{batman:2015} \texttt{celerite} \citep{celerite:2017}, \texttt{dynesty} \citep{dynesty:2020} \texttt{exoplanet} \citep{exoplanet:2021}, \texttt{numpy} \citep{numpy:2020}, \texttt{pandas} \citep{pandas:2010}, \texttt{pyldtk} \citep{pyldtk:2015}, \texttt{pymc3} \citep{pymc3:2016}, \texttt{scipy} \citep{scipy:2020}, \texttt{starry} \citep{starry:2019}.

\end{document}